\documentclass{article}

\pdfoutput=1
\usepackage{fullpage}
\usepackage[utf8]{inputenc}
\usepackage{graphicx}
\usepackage{amssymb}
\usepackage{amsmath}
\usepackage{amsfonts}
\usepackage{color}
\usepackage[pdftex]{hyperref}
\usepackage[english]{babel}
\usepackage{subfigure}
\usepackage{algorithm}
\usepackage{algorithmic}
\usepackage{multicol}
\usepackage{multirow}
\usepackage{tabularx,colortbl}
\usepackage[sort&compress,numbers]{natbib}
\usepackage{authblk}

\definecolor{Orange}{rgb}{1,0.64,0}
\definecolor{lgray}{rgb}{0.9,0.9,0.9}

\makeindex

\title{Analysis of heat kernel highlights the strongly modular and heat-preserving structure of proteins}

\author[1]{Lorenzo Livi\thanks{llivi@scs.ryerson.ca}\thanks{Corresponding author}}
\author[2]{Enrico Maiorino\thanks{enrico.maiorino@uniroma1.it}}
\author[3]{Andrea Pinna\thanks{andrea.pinna@crs4.it}}
\author[1]{Alireza Sadeghian\thanks{asadeghi@ryerson.ca}}
\author[2]{Antonello Rizzi\thanks{antonello.rizzi@uniroma1.it}}
\author[4]{Alessandro Giuliani\thanks{alessandro.giuliani@iss.it}}
\affil[1]{Dept. of Computer Science, Ryerson University, 350 Victoria Street, Toronto, ON M5B 2K3, Canada}
\affil[2]{Dept. of Information Engineering, Electronics, and Telecommunications, SAPIENZA University of Rome, Via Eudossiana 18, 00184 Rome, Italy}
\affil[3]{CRS4 Bioinformatica, Parco Tecnologico della Sardegna, Edificio 1, 09010 Pula (CA), Italy}
\affil[4]{Dept. of Environment and Health, Istituto Superiore di Sanit\`{a}, Viale Regina Elena 299, 00161 Rome, Italy}

\providecommand{\keywords}[1]{\textbf{\textit{Keywords---}} #1}

\begin{document}

\maketitle

\begin{abstract}
In this paper, we study the structure and dynamical properties of protein contact networks with respect to other biological networks, together with simulated archetypal models acting as probes.
We consider both classical topological descriptors, such as the modularity and statistics of the shortest paths, and different interpretations in terms of diffusion provided by the discrete heat kernel, which is elaborated from the normalized graph Laplacians.
A principal component analysis shows high discrimination among the network types, either by considering the topological and heat kernel based vector characterizations. Furthermore, a canonical correlation analysis demonstrates the strong agreement among those two characterizations, providing thus an important justification in terms of interpretability for the heat kernel.
Finally, and most importantly, the focused analysis of the heat kernel provides a way to yield insights on the fact that proteins have to satisfy specific structural design constraints that the other considered networks do not need to obey.
Notably, the heat trace decay of an ensemble of varying-size proteins denotes subdiffusion, a peculiar property of proteins.\\
\end{abstract}

\section{Introduction}

As aptly pointed out by \citet{nicosia2013characteristic} ``Networks are the fabric of complex systems''. Concepts like organized complexity (or middle way \cite{laughlin2000middle}, mesoscopic systems \cite{imry1997introduction}) revolve around the existence of features shared by systems made of many interacting parts, which can be suitably described in terms of networks.
In 1948 \citet{weaver1991science} defined the notions of simplicity and complexity in Science by a three categories classification: (1) \textit{Organized Simplicity}. The paradigm of this kind of problems is exemplified by Newton's law of universal gravitation; (2) \textit{Disorganized Complexity}. In these cases, systems are made by a very large number of interacting elements. Even if these systems cannot be studied at the single-element level, nevertheless they allow for a very efficient statistical treatment (e.g., in terms of statistical mechanics).
There is a third kind of complexity that cannot be studied this way and that Weaver identified with biological systems: \textit{Organized Complexity}.

Both the arguments and terminology used by Weaver were largely coincident with Laughlin \textit{et al.}, identifying ``The Middle Way'' \cite{laughlin2000middle} as the real frontier of basic Science in the XXI century.
The natural entities in which the organized complexity approach was more fruitful were protein molecules.
Not only they were exactly at the boundary between simple and complex physics \cite{frauenfelder1994biomolecules} but they allowed for both very accurate experiments and for a very refined structural description (e.g., via X-ray crystallography).
The situation nowadays continues to be unchanged with respect to Laughlin \textit{et al.} \cite{laughlin2000middle}: protein science is still the most explored field in the realm of organized complexity \cite{vijayabaskar2010interaction,bagler2005network,bode2007network,yan2014construction,doi:10.1021/cr3002356,orozco2014theoretical,banavar2007physics}.

The resurgent interest in graph-theoretical methods for the analysis complex systems \cite{dehmer2013quantitative,ecoli_graph_complexity_arxiv,giuliani2014network,boccaletti+latora+moreno+chavez+hwang2006,csermely2013structure,doi:10.1021/cr3002356,bashan2012network,mirshahvalad2014dynamics,dorogovtsev2008critical,song2006origins,rosvall2014memory,barthelemy2011spatial,holme2012temporal,newman2005power,csermely2012disordered,csermely2013structuredynamics,newman2010networks,song2005self} set forth by the works by Barabasi, Strogatz and other pioneers in the first decade of XXI century, allowed for a rich set of measures providing a multifaceted description of complex networks \cite{Han20121958,newman2005power,ecoli_graph_complexity_arxiv,escolano2012heat,rossi2013characterizing,PhysRevE.80.045102,escolano2012heat,doi:10.1021/ci300321f,duardo2013modeling,riera2012new}.
It is sufficient that a given problem is formalized by means of a vertex-and-edge representation -- with vertices being the parts and edges their pairwise relations -- to allow facing the problem according with the terms of organized complexity \cite{giuliani2014network}.
Protein molecules are at the forefront of complex network way to organized complexity.
In fact, it is possible to represent a protein molecule in terms of protein contact network (i.e., a network having amino acid residues as vertices and edges representing physical contacts between them) by considering its native 3D structure \cite{yan2014construction,estrada2010universality}.
It is worth noting that the graph-based representation is at the core of chemical thought: the structural formulas are graphs \cite{trinajstiac1983chemical}.

In this work, we face a data-driven quest for mesoscopic organization principles of complex biological systems by analyzing different complex networks: protein contact networks, metabolic networks, and genetic networks, together with simulated archetypal networks whose wiring scheme is generated by mathematical rules.
All considered networks are characterized in terms of two collections of numerical features.
The first collection is based on classical topological descriptors, such as the modularity and statistics of the shortest paths. The second one exploits the discrete heat kernel (HK), elaborated using the eigendecomposition of the (normalized) graph Laplacian \cite{Xiao:2009:GCH:1563046.1563099}.
With a first preliminary analysis, we show that the different classes of networks are discriminated by a suitable embedding of the numeric features. This is reasonably expected given the substantially different natures of the analyzed networks, but by no means can be considered as a trivial result.
As a matter of fact, the distinction in terms of metabolic, genetic, and protein contact networks is based on network functions, and the demonstration of a link between functional and structural properties of the corresponding graph representation is a prerequisite for the soundness of the proposed strategy of analysis.
An important result is that the two herein considered network characterizations resulted to be strongly correlated with each other, so giving a proof-of-concept of the reliability and interpretability of the adopted network descriptions.
From this first analysis, it also emerged that protein contact networks display unique properties that do not allow for a straightforward classification in any of the considered classical archetypal networks, stressing the need for the search of new generative models for protein contact networks.
The second, and more important, claim of this paper is that the ensemble HK analysis allows us to derive results in agreement with known chemico-physical properties of protein molecules.
Our analysis demonstrates that a (simulated) diffusion process in protein contact networks proceeds slower than normal diffusion, i.e., we observe subdiffusion. Notably, a two-regime diffusion emerged from the analysis of the heat trace decay: a fast and a slow regime. The fast regime is driven by short cuts putting in contact amino acid residues far-away along the sequence.
Subdiffusion is ubiquitous in Nature \cite{0034-4885-76-4-046602} and it is a well-studied peculiarity describing energy flow \cite{doi:10.1146/annurev.physchem.59.032607.093606,lervik2010heat,li2014anisotropic,sangha2009proteins} and vibration dynamics \cite{yu2003anomalous,reuveni2010anomalies,granek2011proteins,neusius2008subdiffusion,reuveni2012dynamic} in protein molecules.
Interestingly, here we are able to observe the same phenomenon by exploiting only algebraic properties of the considered graph representations of an ensemble of varying-size proteins.
The demonstration of the ability of such a network formalization to explain a unique chemico-physical property of protein molecules, as heat diffusion, is a proof-of-concept of the usefulness of graph theory in chemical systems modeling.

There is sufficient agreement on the fact that proteins, considering their native structure, are highly modular and fractal networks \cite{PhysRevE.79.020901,banerji2011fractal,ecoli_graph_complexity_arxiv,di2012proteins,PhysRevE.71.011912,delvenne2010stability}; yet they are characterized also by short paths connecting distant regions of the molecules responsible for the fast-track transport of energy and protein allosteric properties \cite{doi:10.1146/annurev.physchem.59.032607.093606}.
The modular character of protein contact networks has a direct impact on diffusion processes simulated via graph-theoretical methods.
In fact, well-defined modular organizations slow down diffusion processes in networks \cite{gallos2007scaling}.
Additionally, theoretical and experimental results regarding diffusion on porous and fractal media predict fractional scaling exponents for the mean squared displacement (a well-known measure of diffusion), which give rise to the so-called anomalous diffusion \cite{ben2000diffusion,nakayama1994dynamical}.
To conclude, another interesting fact deduced from our analysis is that, at odds with the other networks, modularity of protein graphs increases with the size of the network.
This observation is fairly interesting since path efficiency and modular properties are two conflicting features in networks \cite{gallos2012small,bullmore2012economy}, which however seem to be suitably optimized in proteins.

The remainder of this paper is organized as follows.
In Sec. \ref{sec:material_and_methods} we first introduce the data that we considered in our study (Sec. \ref{sec:datasets}) and successively we describe the two adopted network characterizations (Sec. \ref{sec:characteristics}).
In Sec. \ref{sec:results} we present the experimental results.
In Sec. \ref{sec:pca} we discuss the results in terms Principal Component Analysis (PCA) of the considered network characterizations in terms of numerical features; in Sec. \ref{sec:cca} we demonstrate the important statistical agreement among the two characterizations.
Finally, in Sec. \ref{sec:scaling} we present the principal results of this paper, in which we show the two-regime diffusion of proteins.
Sec. \ref{sec:conclusions} concludes the paper; Appendix \ref{sec:heat_kernel} provides the HK technical details and Appendix \ref{sec:heat_kernel_spectrum} offers a justification for considering ensemble statistics in the analysis of the HT decay.

\section{Material and Methods}
\label{sec:material_and_methods}

\subsection{The Considered Dataset}
\label{sec:datasets}

Our dataset consists of 323 connected networks (simple graphs).
We considered 100 randomly selected E.Coli protein contact networks (PCN) from the dataset recently elaborated by us \cite{ecoli_graph__arxiv,ecoli_graph_complexity_arxiv} -- in the literature PCN are also referred to as protein contact maps, although here we consider the two denominations as equivalent. Such proteins have been obtained by integrating the Niwa \textit{et al.} \cite{niwa2009} E.Coli data with the available information of the respective native structures gathered from the Protein Data Bank repository \cite{pdb}.
The selected proteins contain from 300 to 1000 amino acid residues. As prescribed by \citet{doi:10.1021/cr3002356}, undirected edges are added among any two alpha carbon atoms within the $[4, 8]$ \AA{} range; the lower 4 \AA{} filter allows to get rid of trivial contacts that do not have important topological properties.
Then, we considered 43 metabolic networks (MN) describing organisms belonging to all three domains of life. Vertices of such networks are the substrates and product of the chemical reactions, while the edges are the reactions catalyzed by the enzymes.
As demonstrated in Ref. \cite{jeong2000large}, those large metabolic networks exhibit a typical scale-free topology.
The size of the 43 metabolic networks ranges from 300 to 1500 vertices.
We also considered a collection of 50 realistic gene regulatory networks (GEN) with a number of vertices varying from 200 to 1100 genes/vertices. The GEN are generated with the SysGenSIM software \cite{pinna2011simulating}, a MATLAB\texttrademark  toolbox for the simulation of systems genetics datasets.
Artificial networks and data by SysGenSIM have already been officially employed for the verification of gene network inference algorithms, such as in the DREAM5 Systems Genetics challenge \cite{dream5}; they have also been used as benchmarks for the development of state-of-the-art reverse-engineering algorithms \cite{pinna2010knockouts}.
The considered GEN networks have been generated with the Exponential Input Power-law Output (EIPO) model, i.e., they are built by (i) sampling the number of ingoing and outgoing edges for each vertex from, respectively, an exponential and a power law distribution, and then by (ii) connecting the vertices accordingly.
These artificial EIPO networks exhibit two well-known structural characteristics of real gene networks: the modularity \cite{barabasi2004network}, and the vertex in-degree and out-degree distributions fitting, respectively, an exponential and a power-law curve \cite{guelzim2002topological}.
Besides the adopted EIPO topology, we considered an average vertex degree varying from 4 to 8: the average degree has been sampled in such a wide range due to the uncertainty in the size of the interactome in typical gene regulatory networks. Apparently, the complexity of biological organisms better correlates with the number of interactions between genes than with the number of genes. Therefore, the average number of edges in gene regulatory networks varies according to the complexity of the represented organism \cite{stumpf2008estimating}; it makes then sense to study gene regulatory networks with a different number of interactions/edges.

To obtain suitable references with the aim of helping us in discussing the results, we considered 130 additional networks of varying size belonging to well-known classes of graphs -- such networks play the role of ``probes'' in our dataset.
Notably, we considered 10 Erd\H{o}s-R\'{e}nyi (ER) graphs generated with probability $p=\log(n)/n$; 10 Barab\'{a}si-Albert (BA) scale-free networks \cite{barabasi1999emergence} with a six-edges preferential attachment scheme; and 10 random regular graphs (REG) with degree equal to six.
To cover all network sizes, such probe networks are generated with a number of vertices ranging from 200 to 1100.
Finally, we also considered the synthetic counterpart of the 100 real proteins (denoted as PCN-S in the following). 
Such synthetic proteins have been generated by considering the same number of vertices and edges of the real proteins.
The generation mechanism of the topology follows the three-rule scheme proposed by \citet{bartoli2007effect}, to simulate the folded configuration of the protein backbone by a probability of contact decreasing with the sequence distance. The only exception is for the rule involving edges of the backbone structure.
In fact, to be consistent with the architecture of our real proteins (we considered edges among residues within 4--8 \AA{}), in PCN-S we added edges only among consecutive residues in the sequence having distance 2. It is worth pointing out that such a generation mechanism gives rise to networks with typical small world topology \cite{bartoli2007effect}.

\subsection{Characterization of the Graph Topology}
\label{sec:characteristics}

In the following, we provide an essential description of the two graph characterizations used in this paper. Full details are omitted for the sake of brevity; we include references to the literature.
In practice, each characterization is meant to offer a description of the original graphs as vectors of numerical features.
The first characterization employs ``classical'' topological descriptors (TD), which include statistics of the degrees/shortest paths and also elaborations of the graph spectrum.
In particular we consider the number of vertices (V) and edges (E) as basic descriptors of the size of the network; the modularity (MOD) \cite{newman2006modularity,blondel2008fast} for quantifying the presence of a global community/cluster structure (please note that we consider as feature the value associated to the partition with maximum modularity); the average closeness centrality (ACC), average shortest path (ASP), average degree centrality (ADC), and average clustering coefficient (ACL) \cite{costa2007characterization}; the energy (EN) and Laplacian energy (LEN) of the spectrum  \cite{gutman2006laplacian}; two invariant features from the heat kernel -- see later for details; the ambiguity (A) \cite{Livi_ga_2013}, which expresses the degree of irregularity of the topology; and finally the entropy of a stationary Markovian random walk (H) \cite{Dehmer201157}.

In the second characterization we exploit the HK only, whose technical details are reported in Appendix \ref{sec:heat_kernel}.
In this respect, we consider here three sets of features: those extracted from the heat trace (HT), the heat content (HC), and the series of invariants (\ref{eq:heat_coeff}) associated to the HC, which are called heat content invariants (HCI).
Please note that HT and HC are time-dependent characteristics, while HCI is not. Therefore, in the characterization exploiting the HK only, we consider several time instants for HT and HC; in the TD characterization, instead, we consider only one time instant for the HT -- corresponding to a ``transient'' regime of the diffusion -- and only the first HCI coefficient. Further details are progressively provided in the following sections.

\section{Results}
\label{sec:results}

The results of the computational experiments are organized in the following subsections. In Sec. \ref{sec:pca} we show and discuss the PCA performed over the two aforementioned characterizations of the considered networks.
In Sec. \ref{sec:cca} we discuss the canonical correlation analysis (CCA) calculated among these two characterizations.
Finally, in Sec. \ref{sec:scaling} we present the results obtained in terms of scaling and diffusion dynamics.

\subsection{Principal Component Analysis}
\label{sec:pca}

Numeric data have been standardized using the component-wise mean and standard deviation.
TD and HK fields were submitted to two separated PCA, projecting hence the networks in the spaces spanned by respective principal components (PC).
The emerging of local models specific for each type of network, and the correlation of the components extracted on the entire data set, is thus a proof of the presence of different architectures characterizing the considered classes of networks.
Moreover, the mutual correlation between PCs of TD and HK -- assessed in Sec. \ref{sec:cca} by a CCA -- gave us a demonstration of the consistency and interpretability of the considered network descriptions.

\subsubsection{Analysis of Topological Descriptors}
\label{sec:pca_td}

\begin{figure*}[ht!]
\centering

\subfigure[PC1--PC2]{
\includegraphics[viewport=0 0 343 245,scale=0.62,keepaspectratio=true]{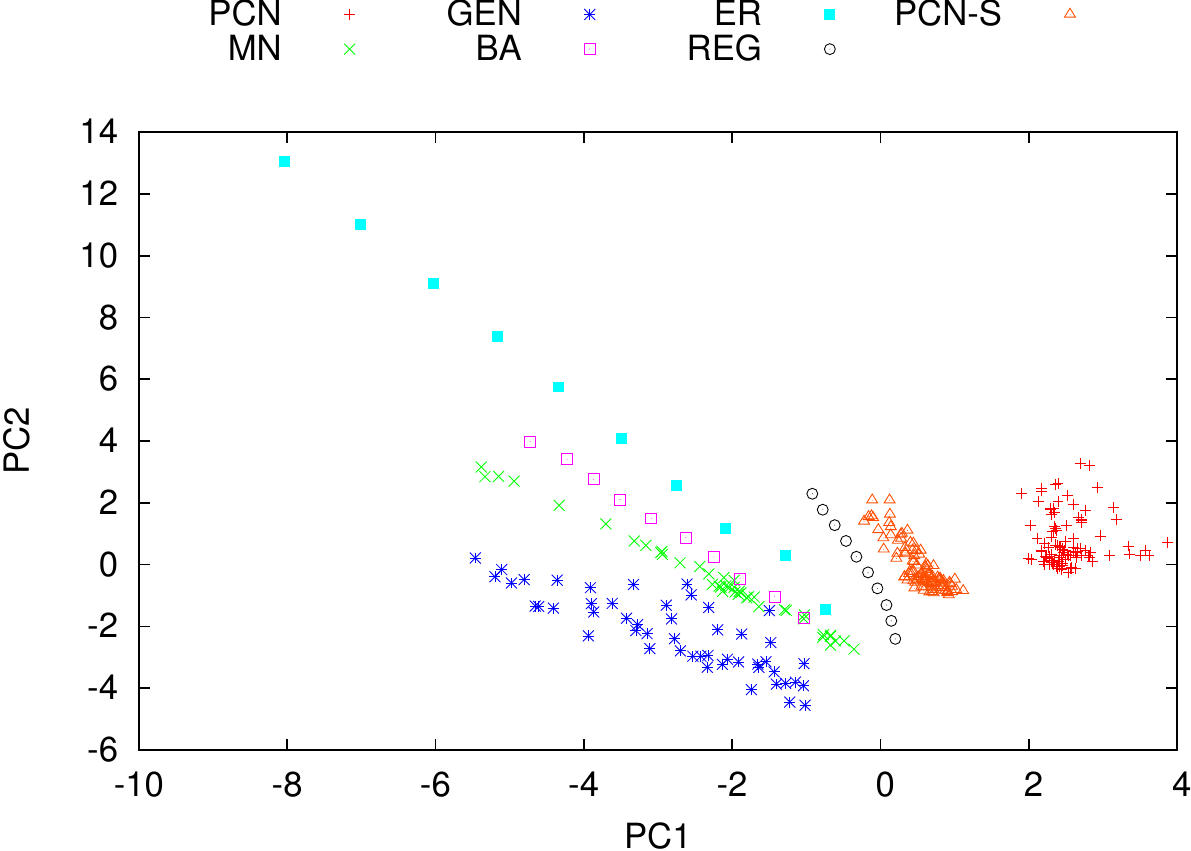}
\label{fig:topo_pc1-2}}
~
\subfigure[PC1--PC3]{
\includegraphics[viewport=0 0 343 245,scale=0.62,keepaspectratio=true]{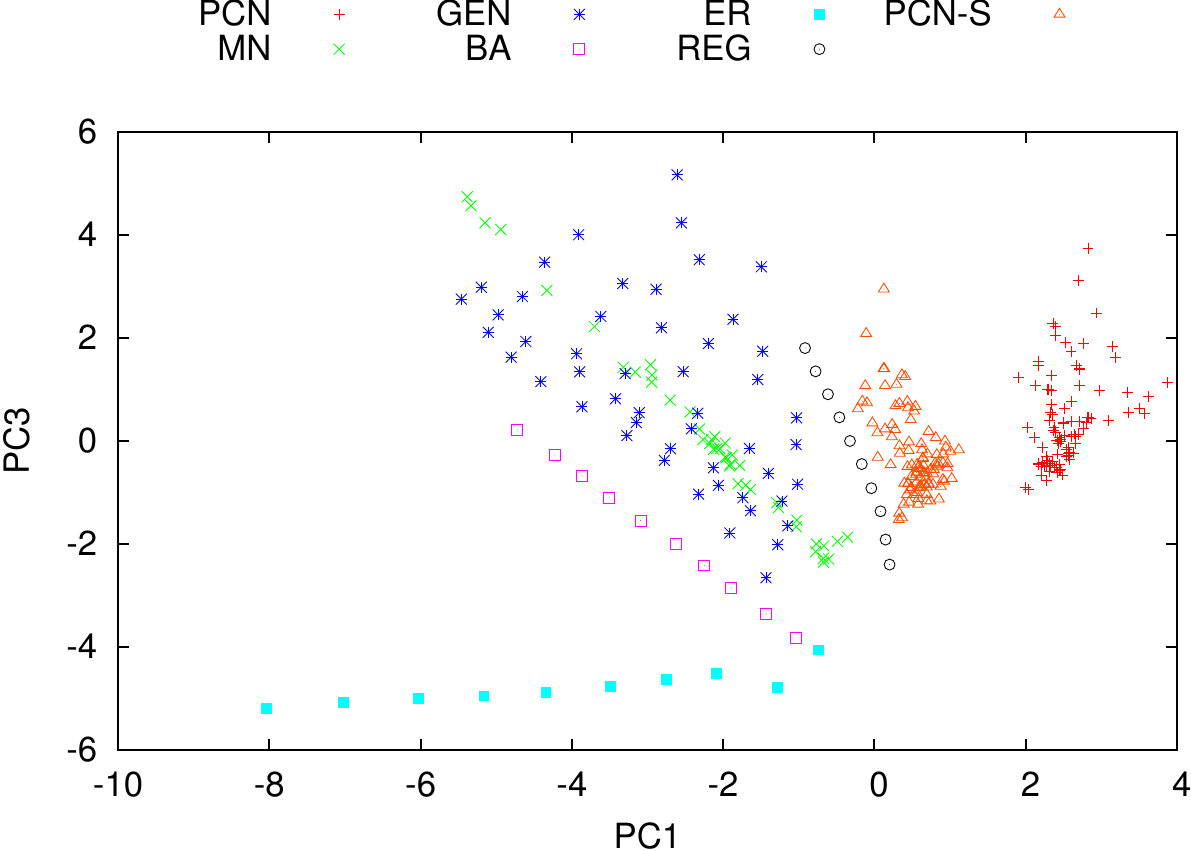}
\label{fig:topo_pc1-3}}

\subfigure[PC2--PC3]{
\includegraphics[viewport=0 0 343 245,scale=0.62,keepaspectratio=true]{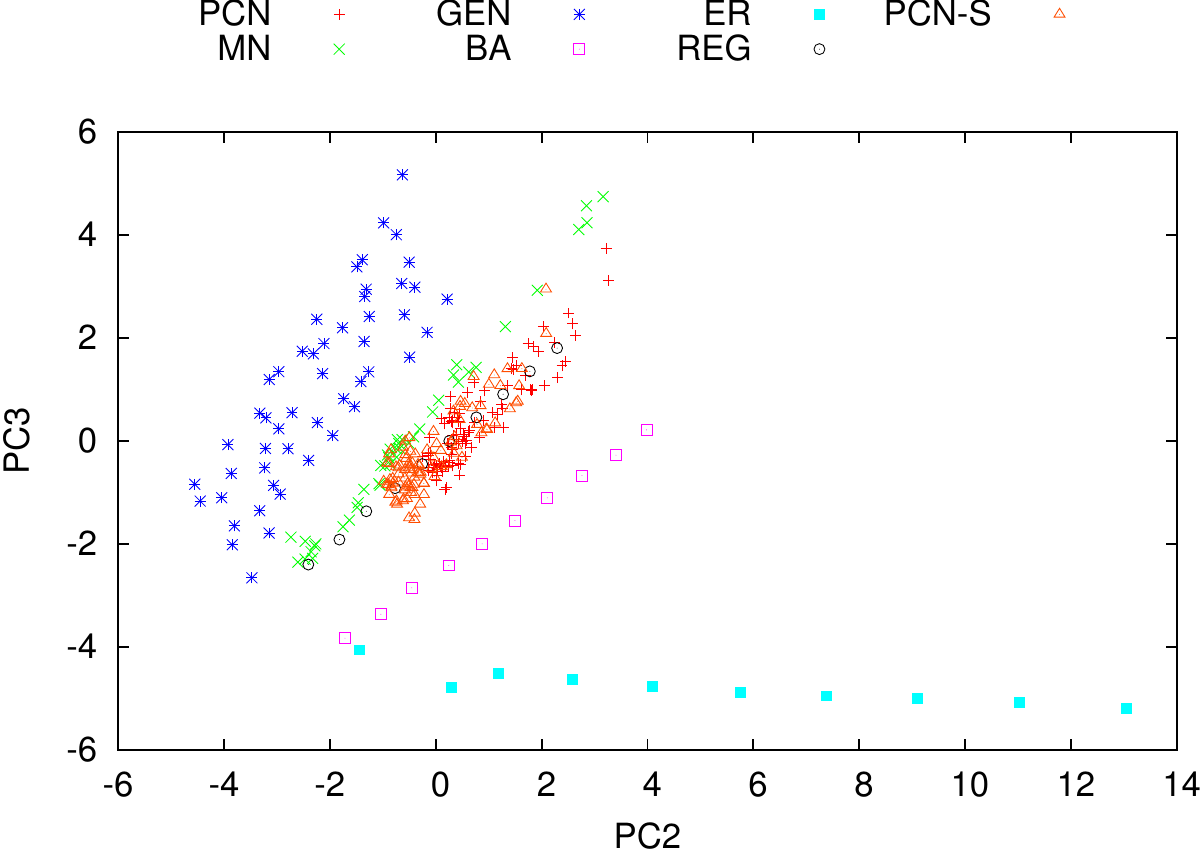}
\label{fig:topo_pc2-3}}
~
\subfigure[PC1--PC2--PC3]{
\includegraphics[viewport=0 0 321 204,scale=0.65,keepaspectratio=true]{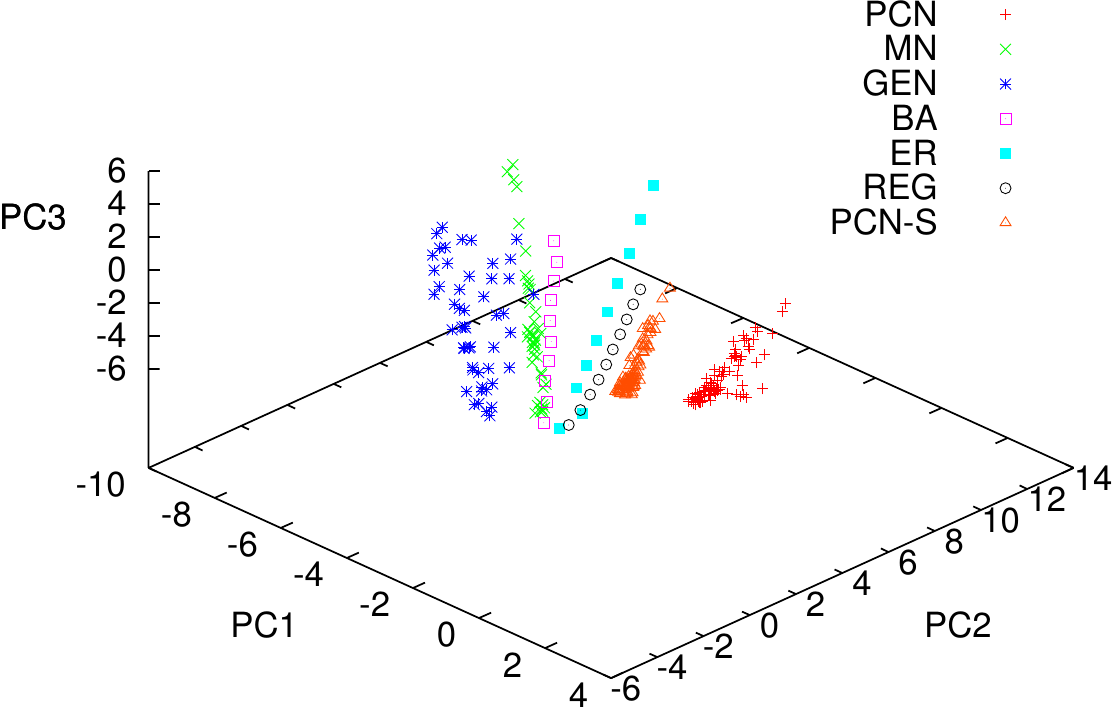}
\label{fig:topo_pc1-2-3}}

\caption{Embedding considering the first three PCs of PCA-TD.}
\label{fig:PCA_Topo_Descriptors}
\end{figure*}

Fig. \ref{fig:PCA_Topo_Descriptors} shows the PCA of the topological descriptors (PCA-TD).
The first three PCs are sufficient to explain more than 90\% of the variance ($\simeq 91\%$).
As it is possible to observe in Fig. \ref{fig:topo_pc1-2}, PC1--PC2 offer a very clear discrimination among the different classes of networks.
The separability persists also by considering PC1--PC3, while however we observe that GEN lose compactness and overlap with MN. By considering PC2--PC3, instead, PCN overlap with REG.
However, the overall picture emerging from PCA-TD clearly points to the possibility of distinguishing among the network types.

Let us now interpret such PCs. Tab. \ref{tab:factors} shows the loadings of the first three factors of PCA-TD.
The first factor (FACTOR-1) is primarily characterized by MOD, ACC, and ASP. As MOD increases (the community structure becomes more evident) preferential paths connecting different regions of the network increase as well. In fact, ACC and ASP are, respectively, negatively and positively correlated with MOD.
It is worth mentioning that ACC and ASP offer a somewhat opposite view of the same feature, i.e., the efficiency of the paths in the networks.
As the global community structure emerges (captured by MOD), also the local clustering structure (ACL) increases as well, although ACL is less loaded on FACTOR-1.
In addition it is worth noting the agreement among the randomness (H) and the modularity: predictability of stationary random walks is affected by the presence of network modules/communities.

FACTOR-2 positively correlates the number of vertices (V) with LEN, which clearly points to the correlation among the network size in terms of number of vertices and the global architecture.
The meaning of this factor will appear more clear in Sec. \ref{sec:scaling}, where we will discuss the scaling of the number of vertices with MOD and the invariant characteristics of the HK.

Finally, the third factor (FACTOR-3) could be interpreted as the ``redundancy'' of the network wiring substrate.
In fact, descriptors heavily loaded on FACTOR-3 are those more directly related to the adjacency matrix--edges.
The ambiguity (A) decreases as the number of edges increases. This means that adding redundancy to the network (i.e., alternative paths) affects the regularity of the topology.

It is immediate to recognize how the different types of networks are characterized by local linear models in the globally orthogonal PC spaces.
These linear models correspond to different scaling relations with network size -- discussed later in Sec. \ref{sec:scaling}.

A simple look at Fig. \ref{fig:PCA_Topo_Descriptors} allows to catch the singular position of PCN on the extreme right of the most informative PC1--PC2 space, so confirming  the peculiar character of PCN  with respect to classical network architectures. Moreover it is worth noting that the artificial polymer networks -- PCN-S -- are the most similar to PCN, although it is not possible to appreciate any overlap. This fact suggests that proteins are not just ``coiled strings'' as hypothesized in Ref. \cite{bartoli2007effect}.
In addition to the features coming from the folding of a continuous backbone, PCN have other peculiar characteristics.
\begin{table}[thp!]\scriptsize
\begin{center}
\caption{Loadings of the first three factors of PCA-TD. Relevant correlations are in bold.}
\label{tab:factors}
\begin{tabular}{|c|c|c|c|}
\hline
\textbf{DESCRIPTOR} & \textbf{FACTOR-1} & \textbf{FACTOR-2} & \textbf{FACTOR-3} \\
\hline
V & -0.0441 & \textbf{0.9953} & 0.0513 \\
E & -0.2053 & 0.5095 & \textbf{0.8082} \\
MOD & \textbf{0.9591} & -0.1383 & -0.1036 \\
ADC & -0.2294 & 0.0428 & \textbf{0.9375} \\
ACC & \textbf{-0.9918} & 0.0353 & 0.0637 \\
ASP & \textbf{0.9281} & -0.0279 & 0.0315 \\
ACL & 0.6716 & -0.3588 & -0.0010 \\
EN & -0.0166 & 0.6830 & \textbf{0.7268} \\
LEN & -0.3944 & \textbf{0.8407} & 0.1712 \\
HT ($t=5$) & 0.6696 & 0.6486 & -0.1007 \\
HCI ($m=1$) & 0.4914 & -0.6172 & 0.4639 \\
H & 0.6906 & -0.2774 & 0.4918 \\
A & -0.3229 & 0.1878 & \textbf{-0.7584} \\
\hline
\end{tabular}
\end{center}
\end{table}

\subsubsection{Analysis of the Heat Kernel}
\label{sec:anal_hk}

We consider three types of invariant features elaborated from the HK: HT, HC, and HCI.
For the PCA of HT and HC we take into account ten time instants going from $t=0$ to $t=9$; this choice will be justified later in Sec. \ref{sec:scaling}; for the PCA of HCI we consider the first ten coefficients $q_m$ of the series in Eq. \ref{eq:heat_coeff} -- this choice is motivated by the fact that for higher-order coefficients the values become numerically unstable.
In all cases, the first three PCs are sufficient to explain more than 90\% of the variance of the original data, and so they are retained for the embedding.

\begin{figure*}[ht!]
\centering

\subfigure[PC1--PC2.]{
\includegraphics[viewport=0 0 343 245,scale=0.62,keepaspectratio=true]{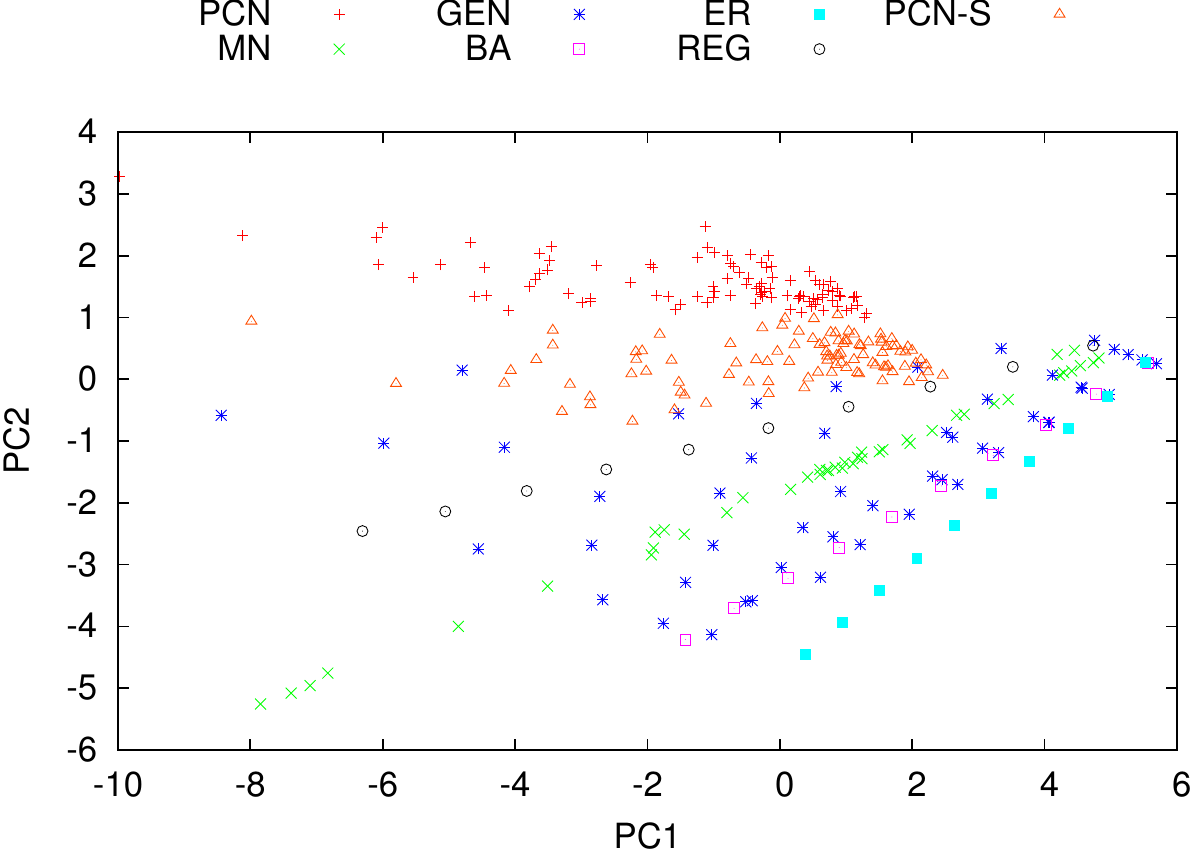}
\label{fig:HT_PC1-PC2}}
~
\subfigure[PC1--PC3.]{
\includegraphics[viewport=0 0 343 245,scale=0.62,keepaspectratio=true]{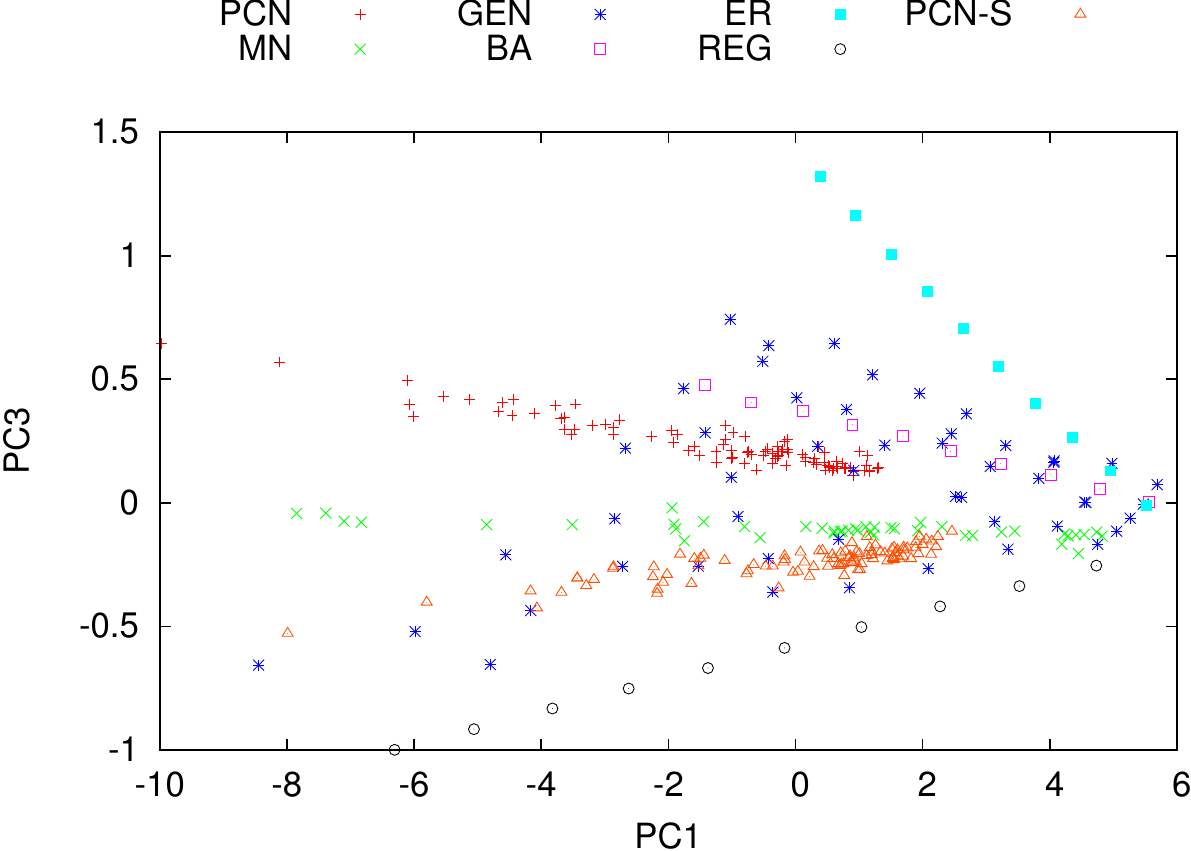}
\label{fig:HT_PC1-PC3}}

\subfigure[PC2--PC3.]{
\includegraphics[viewport=0 0 343 245,scale=0.62,keepaspectratio=true]{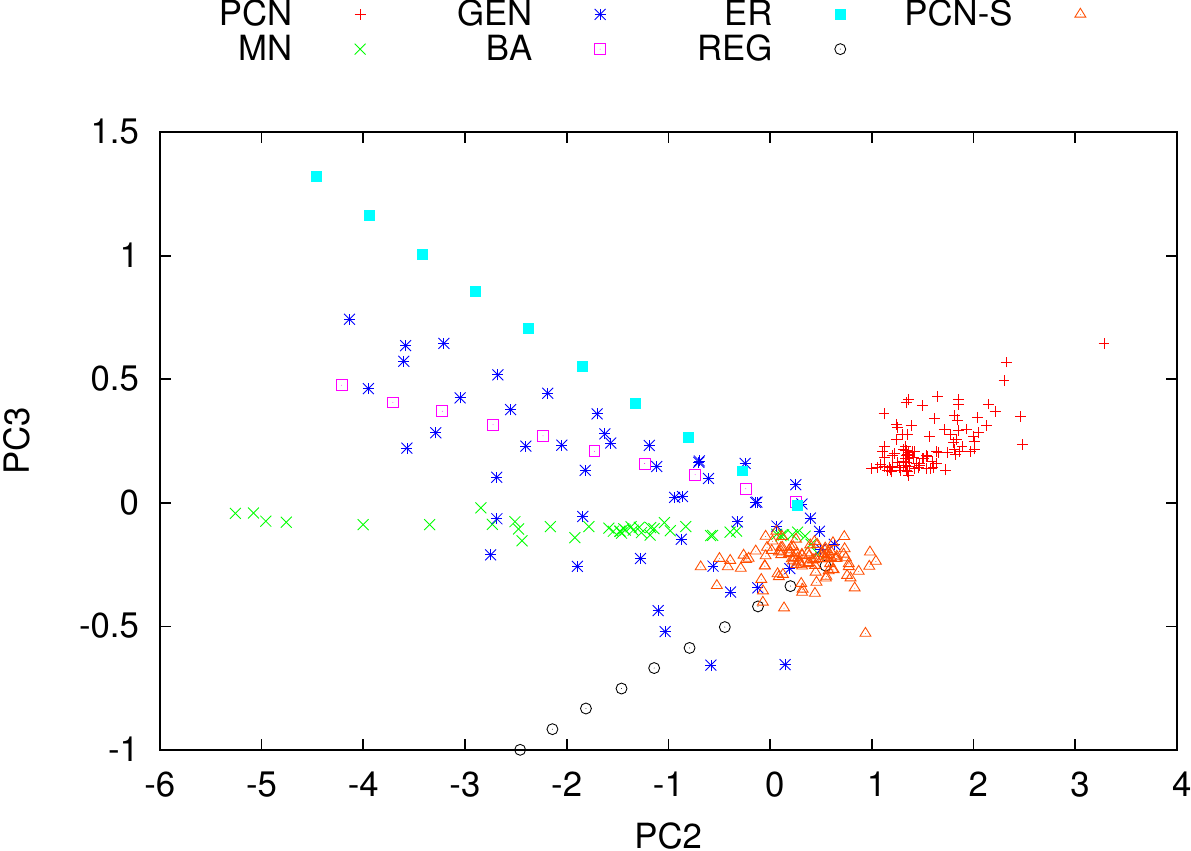}
\label{fig:HT_PC2-PC3}}
~
\subfigure[PC1--PC2--PC3.]{
\includegraphics[viewport=0 0 318 195,scale=0.65,keepaspectratio=true]{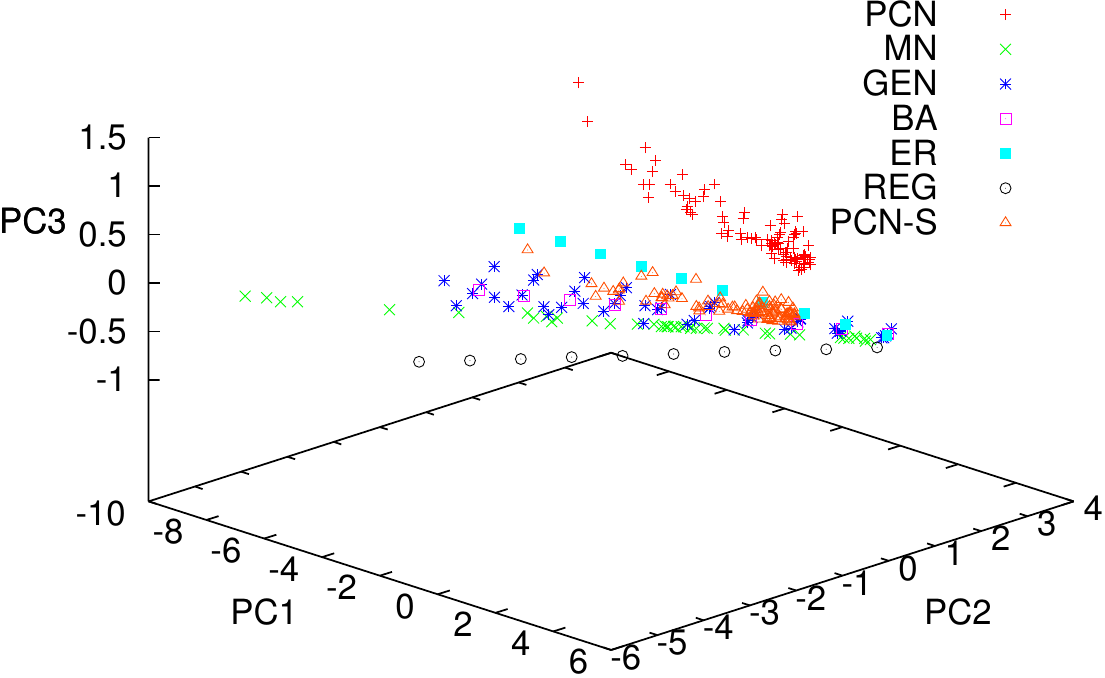}
\label{fig:HT_PC1-PC2-PC3}}

\caption{Embedding of the first three PCs of PCA-HT considering $t=0, 1, ..., 9$.}
\label{fig:PCA_HT}
\end{figure*}

\begin{figure*}[ht!]
\centering

\subfigure[PC1--PC2.]{
\includegraphics[viewport=0 0 343 245,scale=0.62,keepaspectratio=true]{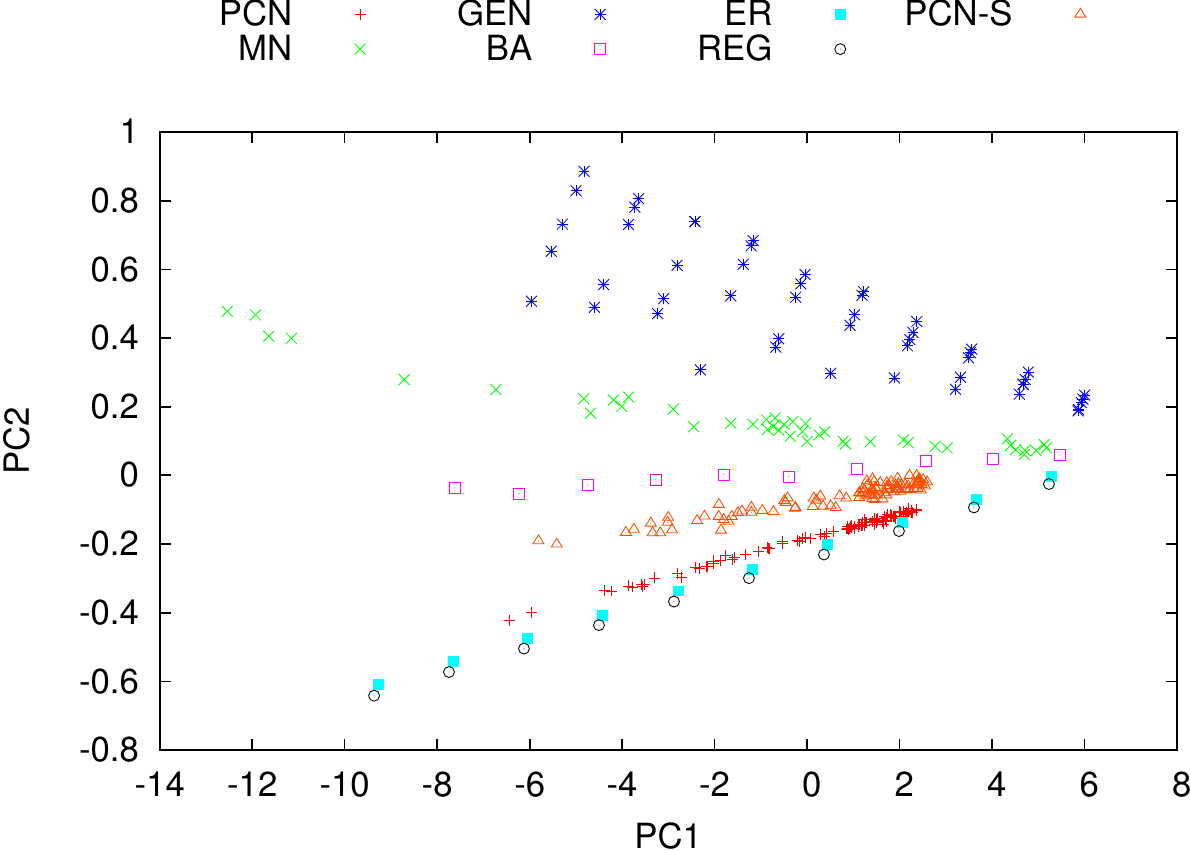}
\label{fig:HC_PC1-PC2}}
~
\subfigure[PC1--PC3.]{
\includegraphics[viewport=0 0 343 245,scale=0.62,keepaspectratio=true]{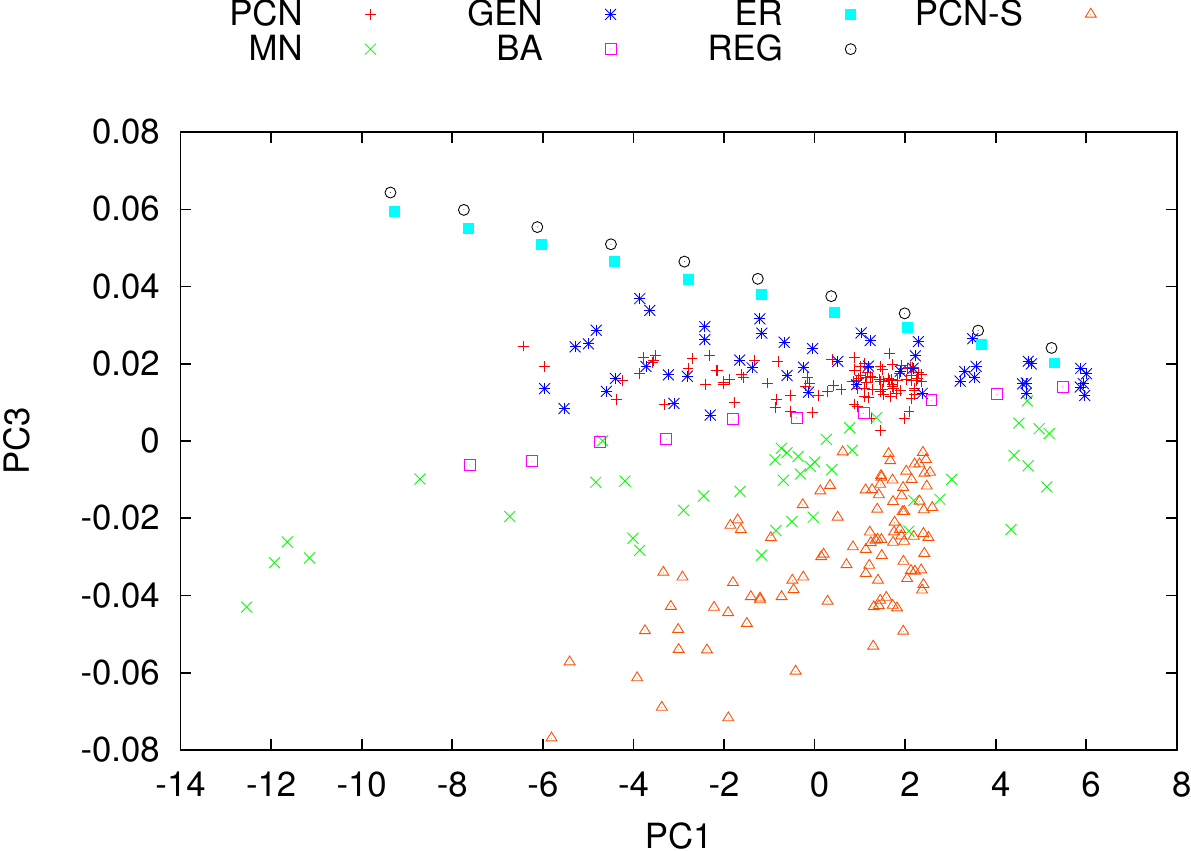}
\label{fig:HC_PC1-PC3}}

\subfigure[PC2--PC3.]{
\includegraphics[viewport=0 0 343 245,scale=0.62,keepaspectratio=true]{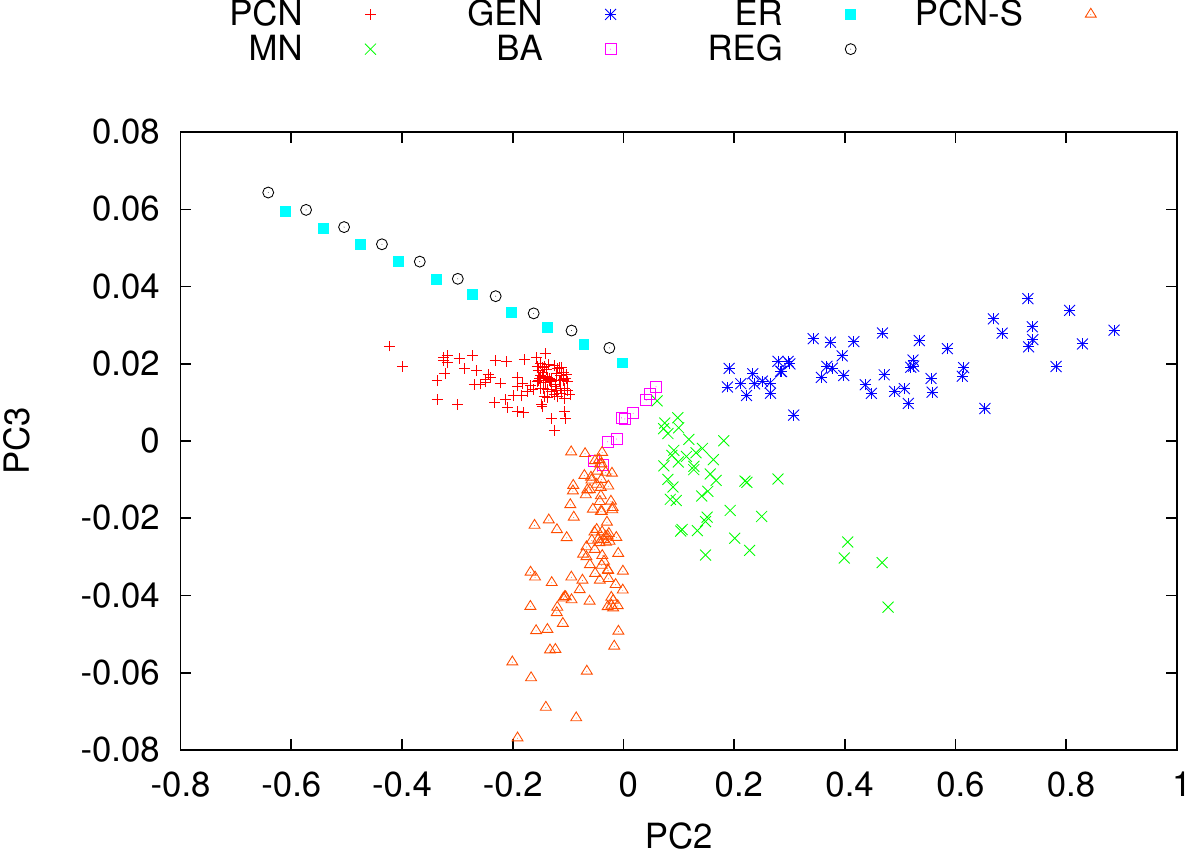}
\label{fig:HC_PC2-PC3}}
~
\subfigure[PC1--PC2--PC3.]{
\includegraphics[viewport=0 0 318 199,scale=0.65,keepaspectratio=true]{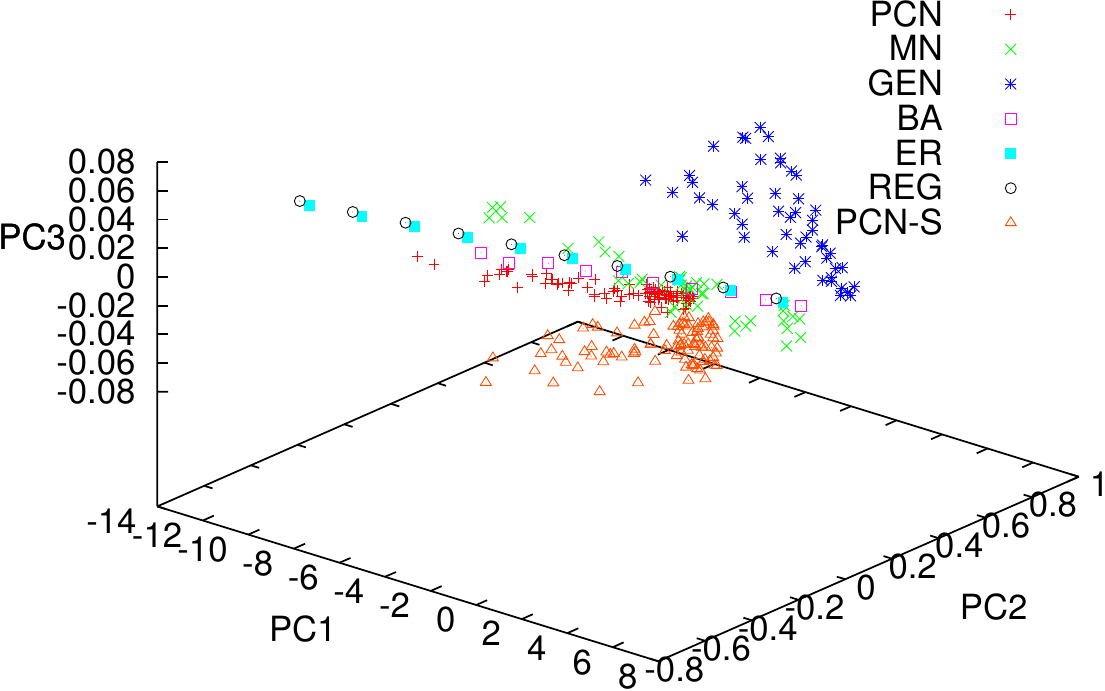}
\label{fig:HC_PC1-PC2-PC3}}

\caption{Embedding of the first three PCs of PCA-HC considering $t=0, 1, ..., 9$.}
\label{fig:PCA_HC}
\end{figure*}

In Fig. \ref{fig:PCA_HT} it is shown the PCA of the HT representation (PCA-HT).
From PC1--PC2 and PC2--PC3 of PCA-HT it is possible to understand that PCN are clearly recognizable by considering the HT, while however GEN depict a not-so-coherent pattern (this is valid for all three PCs).
In Fig. \ref{fig:PCA_HC}, instead, we show the PCA of the HC representation (PCA-HC). We remind to the reader that HT and HC are correlated, since HC considers the information provided by both eigenvalues and eigenvectors of the normalized Laplacian, and not just the eigenvalues as in the HT case.
In fact, from PCA-HC it is possible to note that all networks denote a clear distinguishability; considering either PC1--PC2 and PC2--PC3 almost all networks seems to denote a very peculiar configuration in the PCA space.

\begin{figure*}[ht!]
\centering

\subfigure[PC1--PC2.]{
\includegraphics[viewport=0 0 346 245,scale=0.62,keepaspectratio=true]{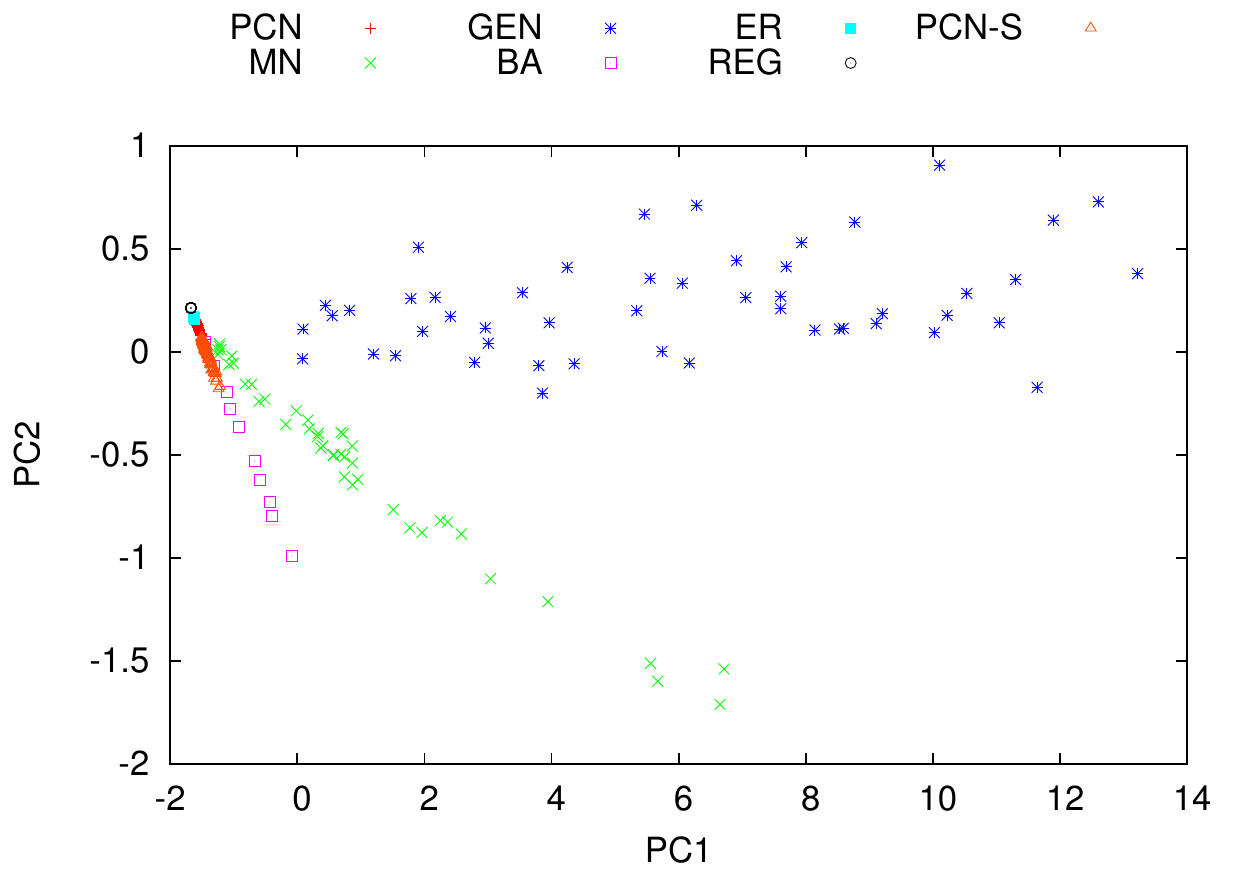}
\label{fig:HCI_PC1-PC2}}
~
\subfigure[PC1--PC3.]{
\includegraphics[viewport=0 0 346 245,scale=0.62,keepaspectratio=true]{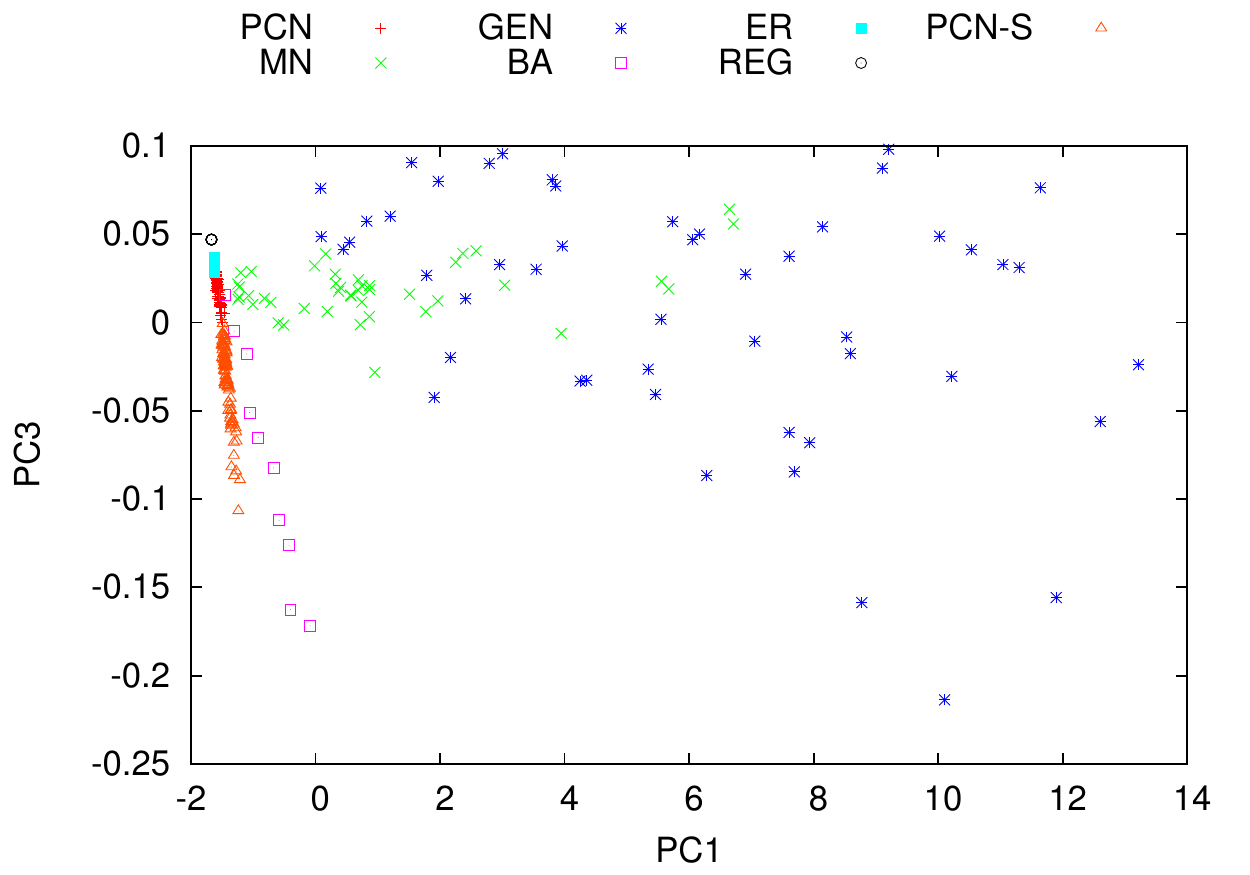}
\label{fig:HCI_PC1-PC3}}

\subfigure[PC2--PC3.]{
\includegraphics[viewport=0 0 346 245,scale=0.62,keepaspectratio=true]{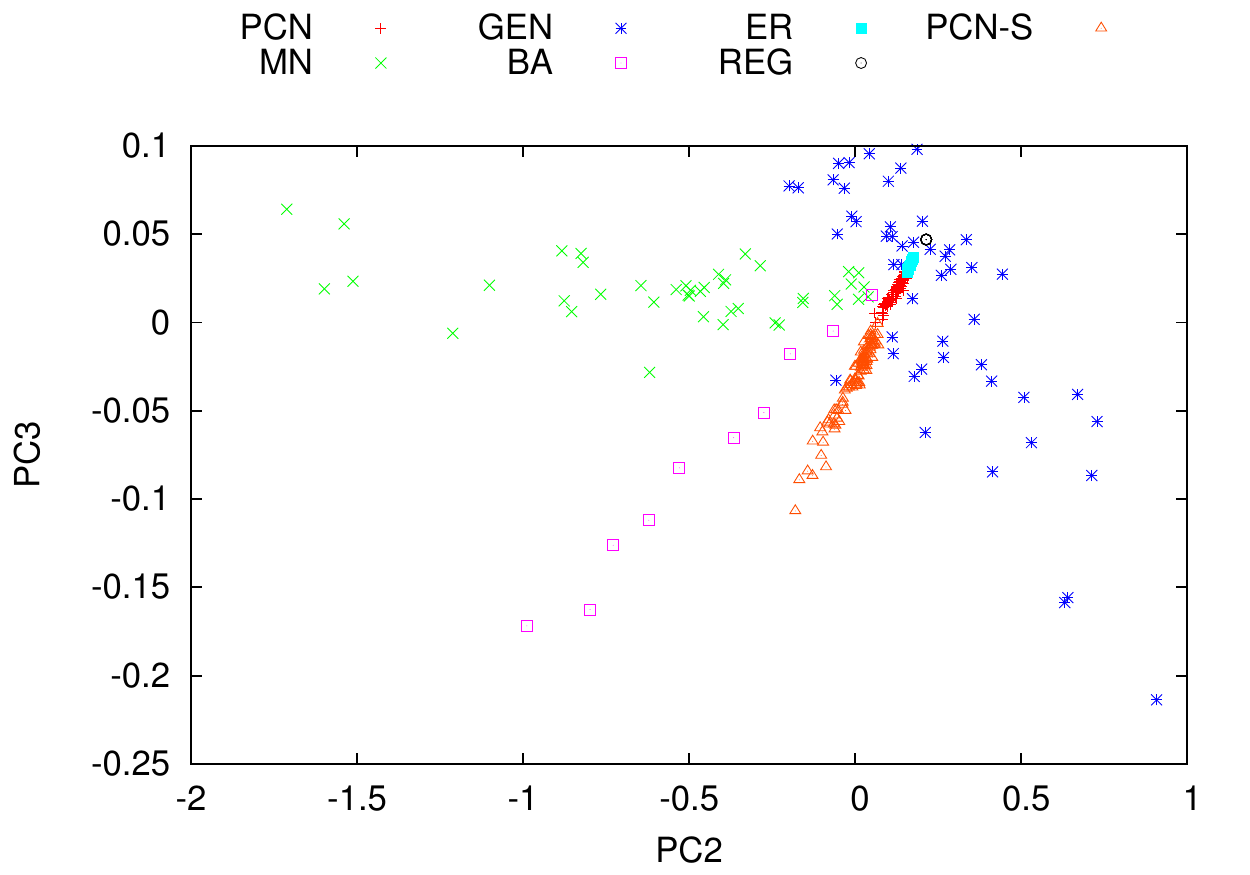}
\label{fig:HCI_PC2-PC3}}
~
\subfigure[PC1--PC2--PC3.]{
\includegraphics[viewport=0 0 318 195,scale=0.65,keepaspectratio=true]{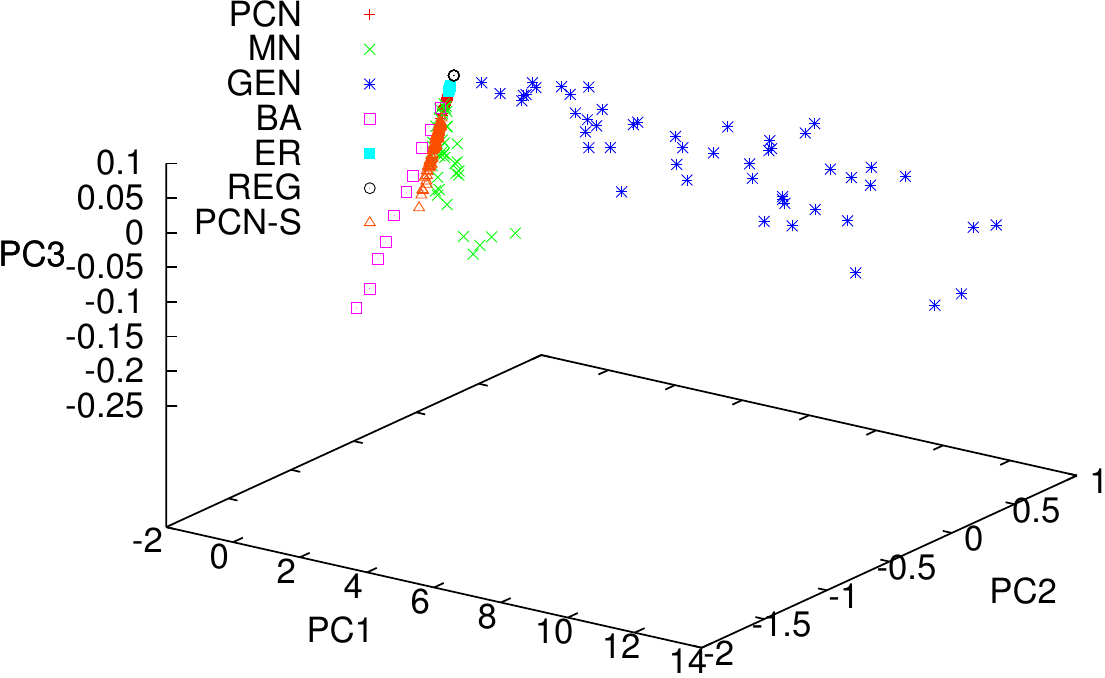}
\label{fig:HCI_PC1-PC2-PC3}}

\caption{Embedding of the first three PCs of PCA-HCI considering the first ten HCI coefficients.}
\label{fig:PCA_HCI}
\end{figure*}

Finally, in Fig. \ref{fig:PCA_HCI} we show the PCA of the HCI representation (PCA-HCI).
The PCs of PCA-HCI allow us noting how PCN, REG, and ER denote a very compact configuration in the PCA-HCI space, while GEN, BA, and MN present a more sparse distribution.
This fact might be interpreted by observing that such two groups differentiate among networks having a clear scale-free topology (second group) and those that are not scale-free (first group).
Interestingly, PCN-S seem to lie in-between those two groups.

\subsection{Canonical Correlation Analysis of the PCA Representations}
\label{sec:cca}

Here we discuss the canonical correlation analysis (CCA) calculated among the various PCA discussed in Sec. \ref{sec:pca}.
For the CCA, we always consider the first three PCs of each representation.
In Tab. \ref{tab:cca} are reported the pairwise correlation values among the most important canonical variates.
There is a strong agreement among all considered PCA representations.
Since part of the information from the HK is present also in TD, we have considered also a PCA representation of TD that does not include such information -- in the table it is indicated as ``PCA-TD\_NO-HK''.
Interestingly, removing the information of the HK from the TD does not alter the scored correlation, so giving a demonstration of the strong coherence between TD and HK based representations of the considered networks.
This result suggests the possibility to interpret the three HK based representations in terms of the more understandable linear correlation structure discussed in Sec. \ref{sec:pca_td}.
\begin{table}[thp!]\scriptsize
\begin{center}
\caption{Canonical correlation coefficients between the first canonical variates relative to different principal component spaces.}
\label{tab:cca}
\begin{tabular}{|c|c|c|c|}
\hline
& \textbf{PCA-HT} & \textbf{PCA-HC} & \textbf{PCA-HCI} \\
\hline
\textbf{PCA-TD} & 0.993 & 0.992 & 0.961 \\
\textbf{PCA-TD\_NO-HK} & 0.988 & 0.986 & 0.946 \\
\hline
\end{tabular}
\end{center}
\end{table}

\subsection{Scaling and Heat Diffusion Analysis}
\label{sec:scaling}

In this section we first study the results in terms of scaling of MOD, HT, HC, and HCI w.r.t. the number of vertices of the considered networks.
Successively, we provide an analysis of the characteristic diffusion patterns emerged from the focused study of the HK.

\begin{figure*}[ht!]
\centering

\subfigure[Scaling of modularity.]{
\includegraphics[viewport=0 0 350 245,scale=0.62,keepaspectratio=true]{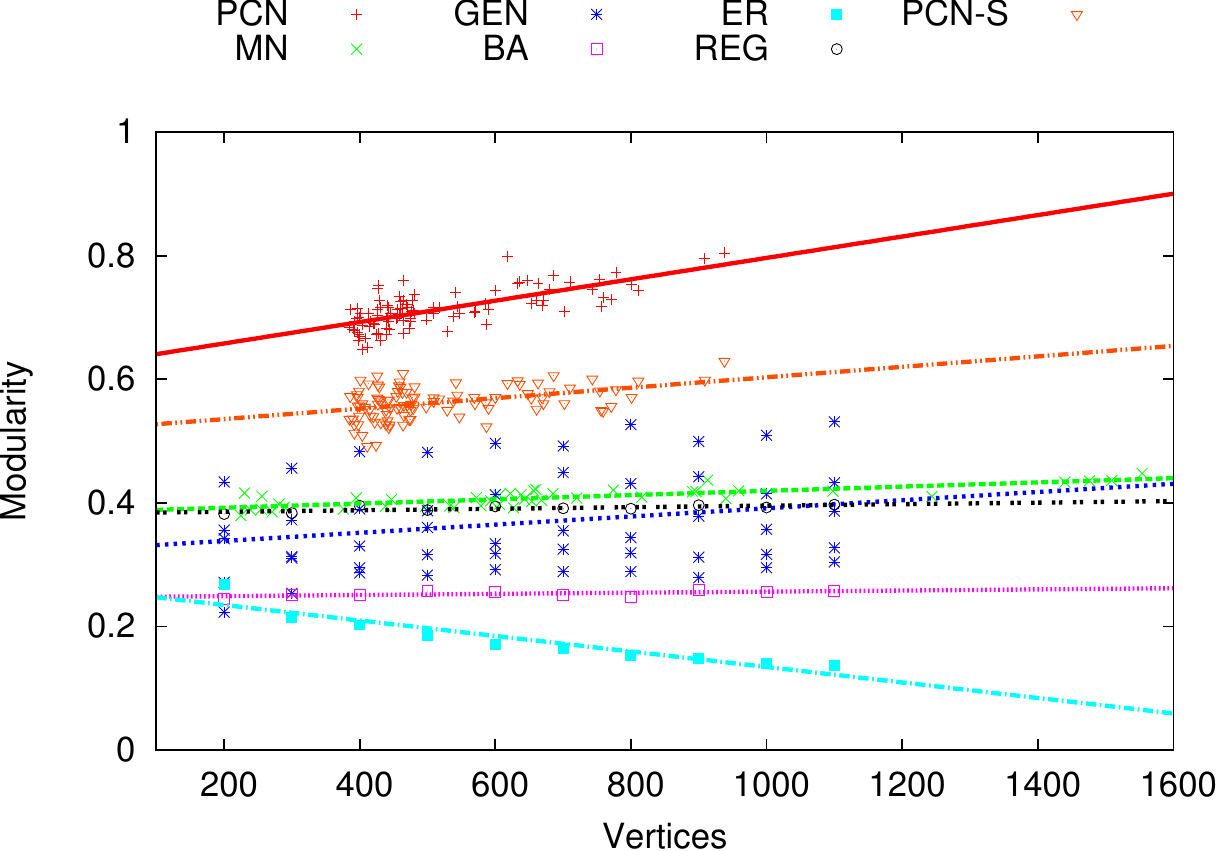}
\label{fig:scaling_modularity_all}}
~
\subfigure[Modularity of GEN with different average degree.]{
\includegraphics[viewport=0 0 339 245,scale=0.62,keepaspectratio=true]{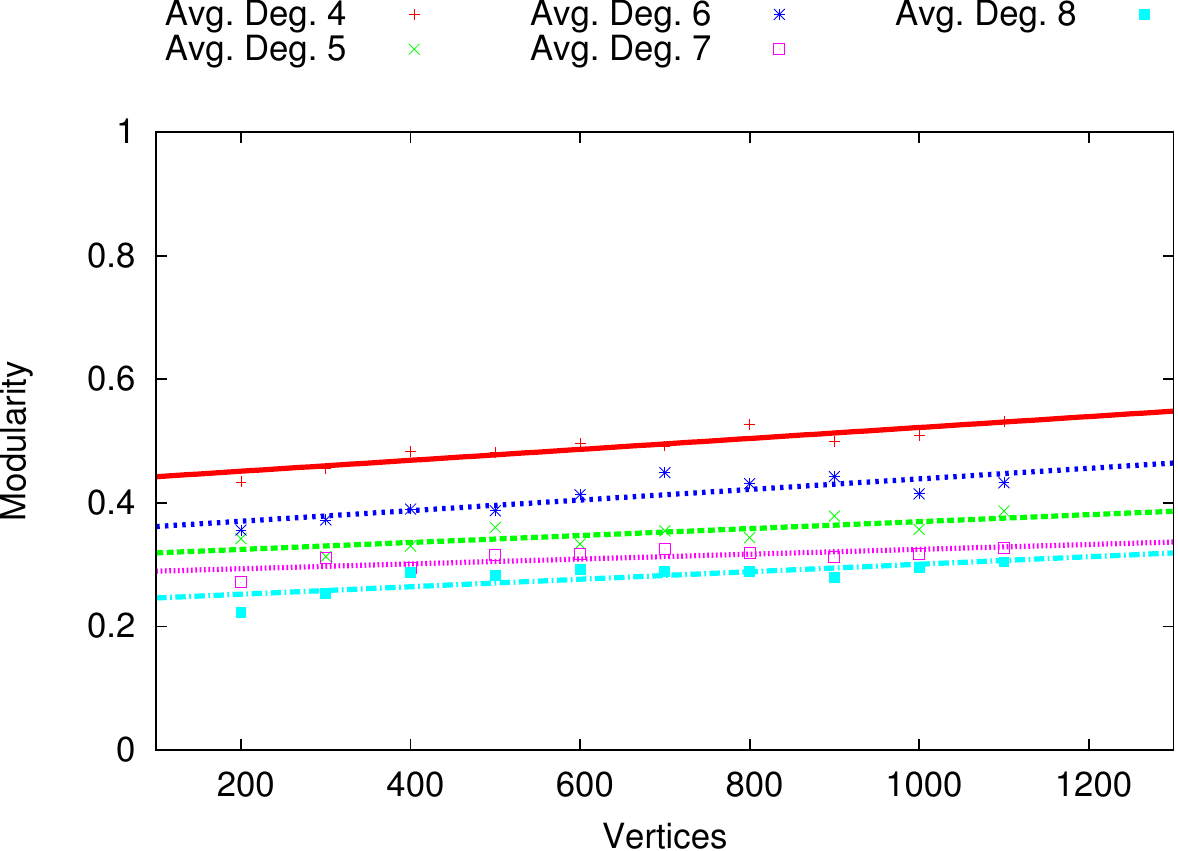}
\label{fig:scaling_modularity_ge}}

\caption{Scaling of modularity over network size. Please note that the linear fitting lines are introduced only to visually guide the reader, and are not meant to provide a model describing the MOD trend asymptotically.}
\label{fig:scaling_modularity}
\end{figure*}

Fig. \ref{fig:scaling_modularity} shows the scaling of the modularity with the size (number of vertices) of the networks.
As already noted in Tab. \ref{tab:factors}, V and MOD do not appear to be globally correlated.
In fact, PCN and PCN-S are the only architectures that show an increasing trend, while the others appear to be almost uncorrelated. We note an exception for ER that tend toward a negative correlation; please note that in the ER case, analytical results are available \cite{guimera2004modularity}.
It is worth explaining the particular pattern of GEN, which does not show a clear trend. In Fig. \ref{fig:scaling_modularity_ge} we show the scaling for GEN by considering the different average degrees used for the EIPO model, where we can observe that each average degree gives rise to a definite trend of MOD.
Please note that the linear fitting lines in Fig. \ref{fig:scaling_modularity} are introduced only to help the reader visually, and are not meant to provide a model describing the MOD trend asymptotically.

\begin{figure*}[ht!]
\centering

\subfigure[Scaling of HT for $t=1$.]{
\includegraphics[viewport=0 0 350 245,scale=0.62,keepaspectratio=true]{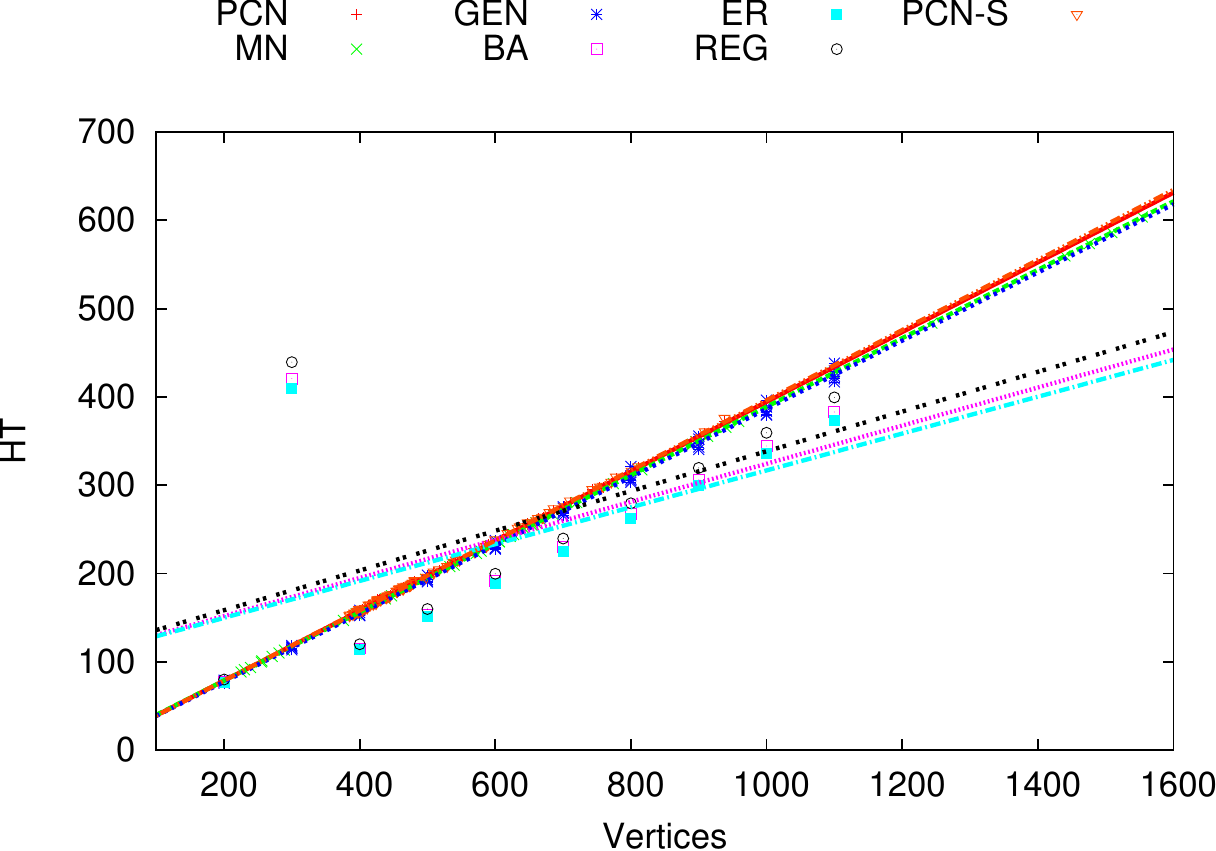}
\label{fig:scaling_HT_t1}}
~
\subfigure[Scaling of HT for $t=5$.]{
\includegraphics[viewport=0 0 350 245,scale=0.62,keepaspectratio=true]{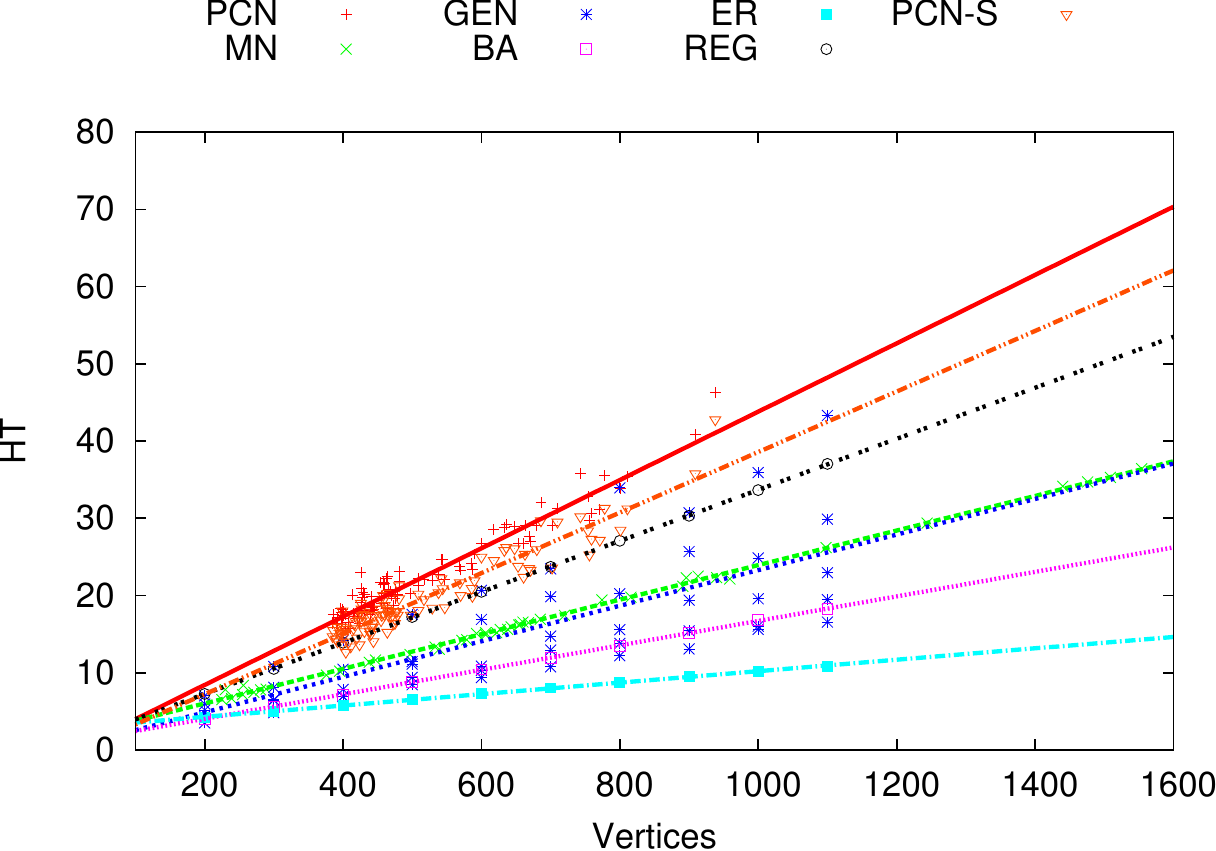}
\label{fig:scaling_HT_t5}}

\subfigure[Scaling of HT for $t=9$.]{
\includegraphics[viewport=0 0 350 245,scale=0.62,keepaspectratio=true]{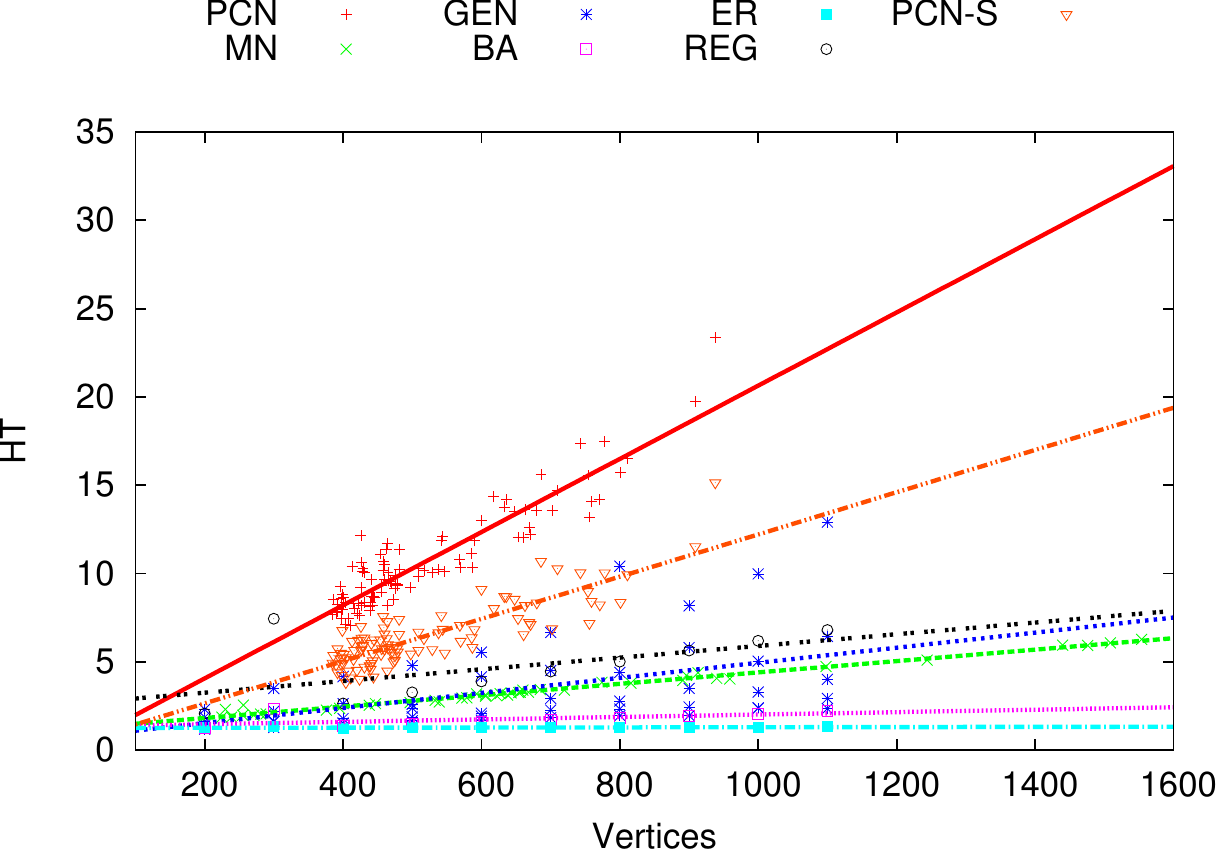}
\label{fig:scaling_HT_t9}}
~
\subfigure[Linear correlations among MOD and $\mathrm{HT}(t)$.]{
\includegraphics[viewport=0 0 349 241,scale=0.62,keepaspectratio=true]{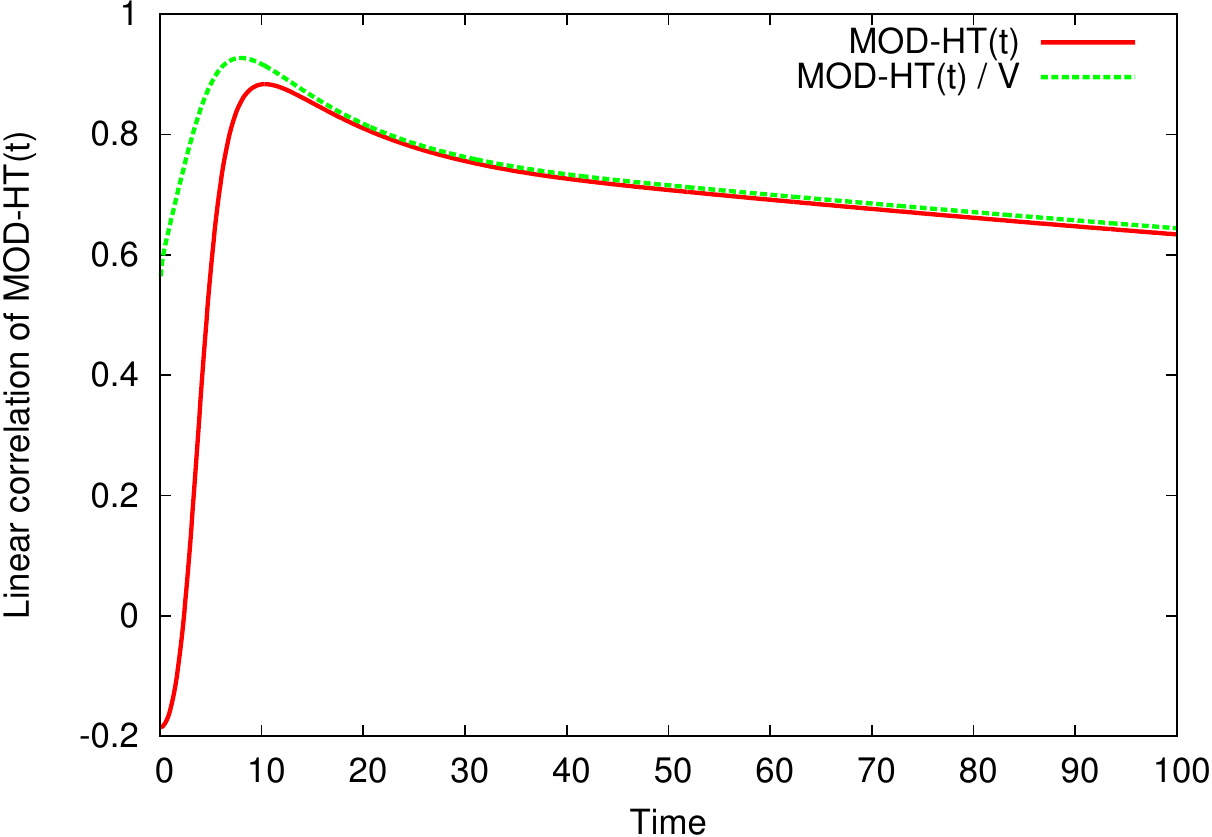}
\label{fig:correlations}}

\caption{Scaling of HT over networks size.}
\label{fig:scaling_HT}
\end{figure*}

\begin{figure*}[ht!]
\centering

\subfigure[Linear fitting slopes of HT over time.]{
\includegraphics[viewport=0 0 340 247,scale=0.6,keepaspectratio=true]{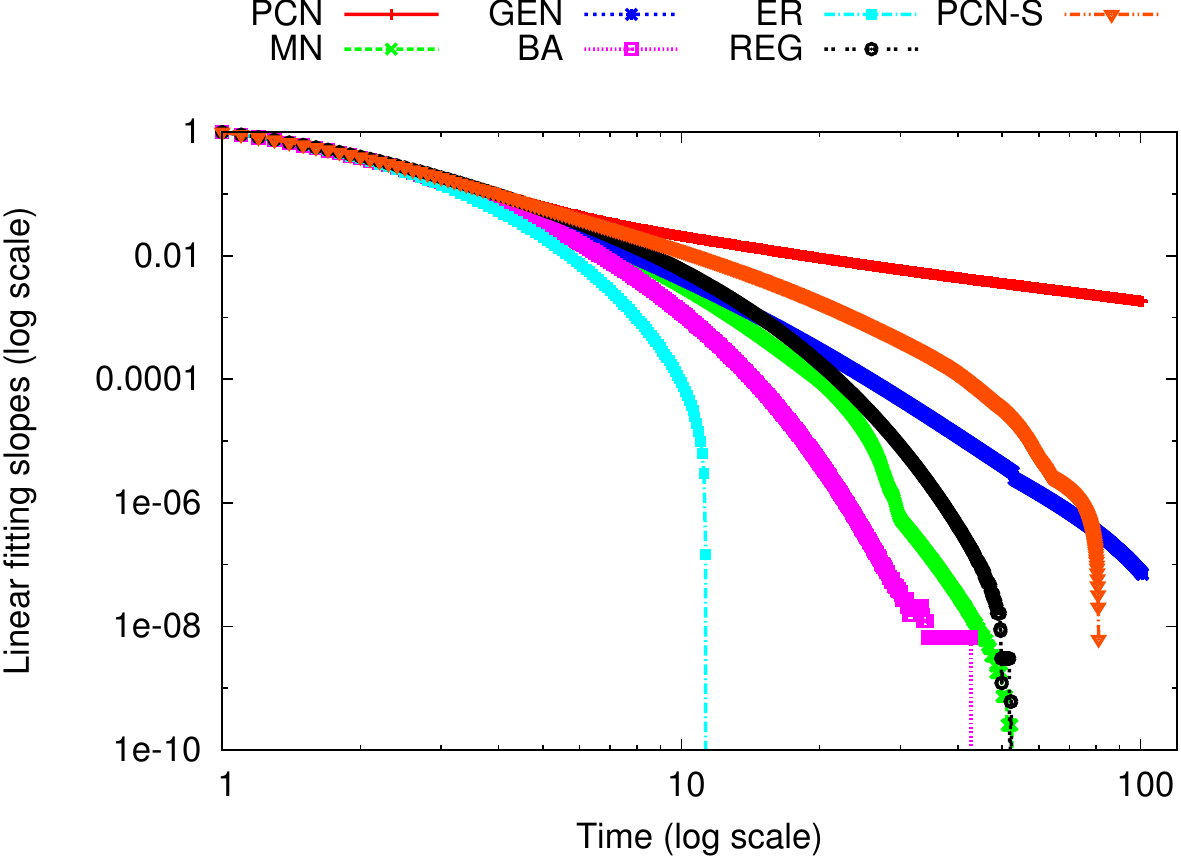}
\label{fig:scaling_HT_slopes_time}}
~
\subfigure[Linear fitting slopes of HT over time (PCN only).]{
\includegraphics[viewport=0 0 340 247,scale=0.6,keepaspectratio=true]{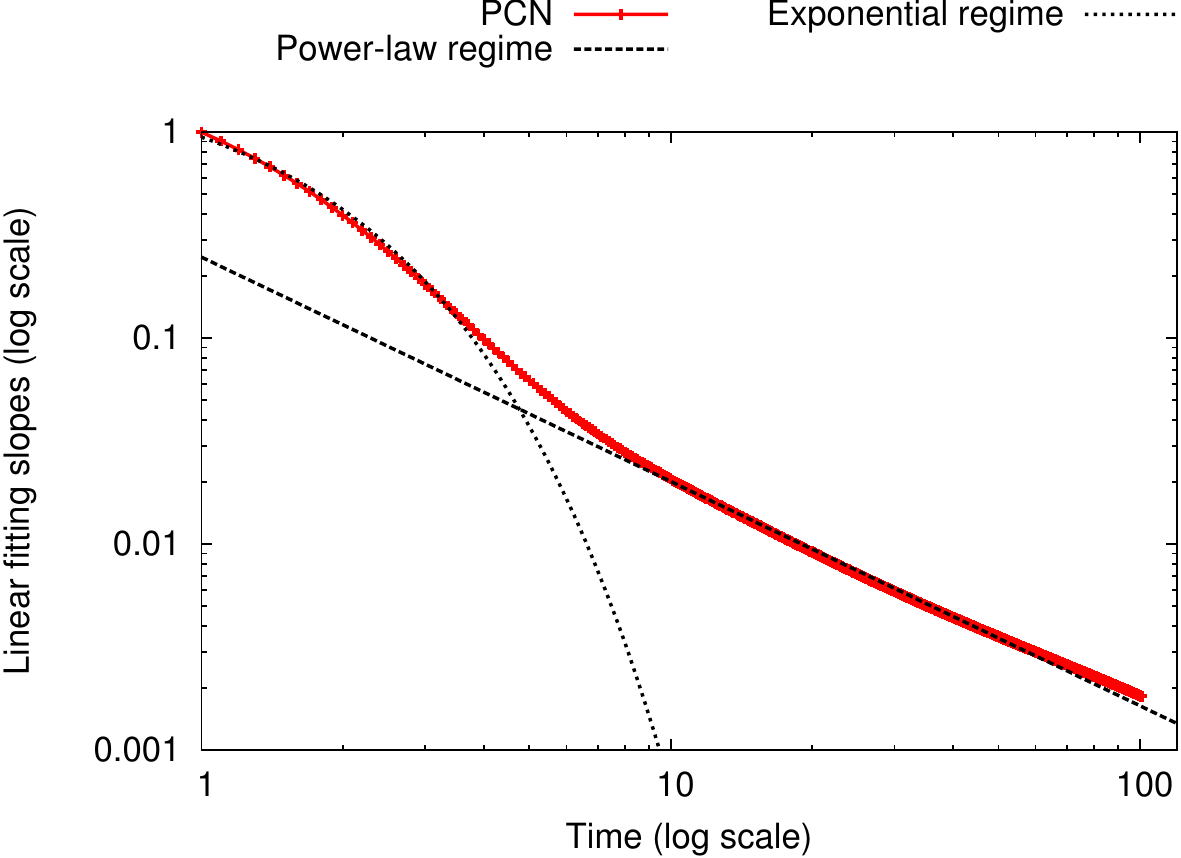}
\label{fig:scaling_HT_slopes_time__P}}

\subfigure[Linear fitting slopes of HT over time (MN only).]{
\includegraphics[viewport=0 0 340 247,scale=0.6,keepaspectratio=true]{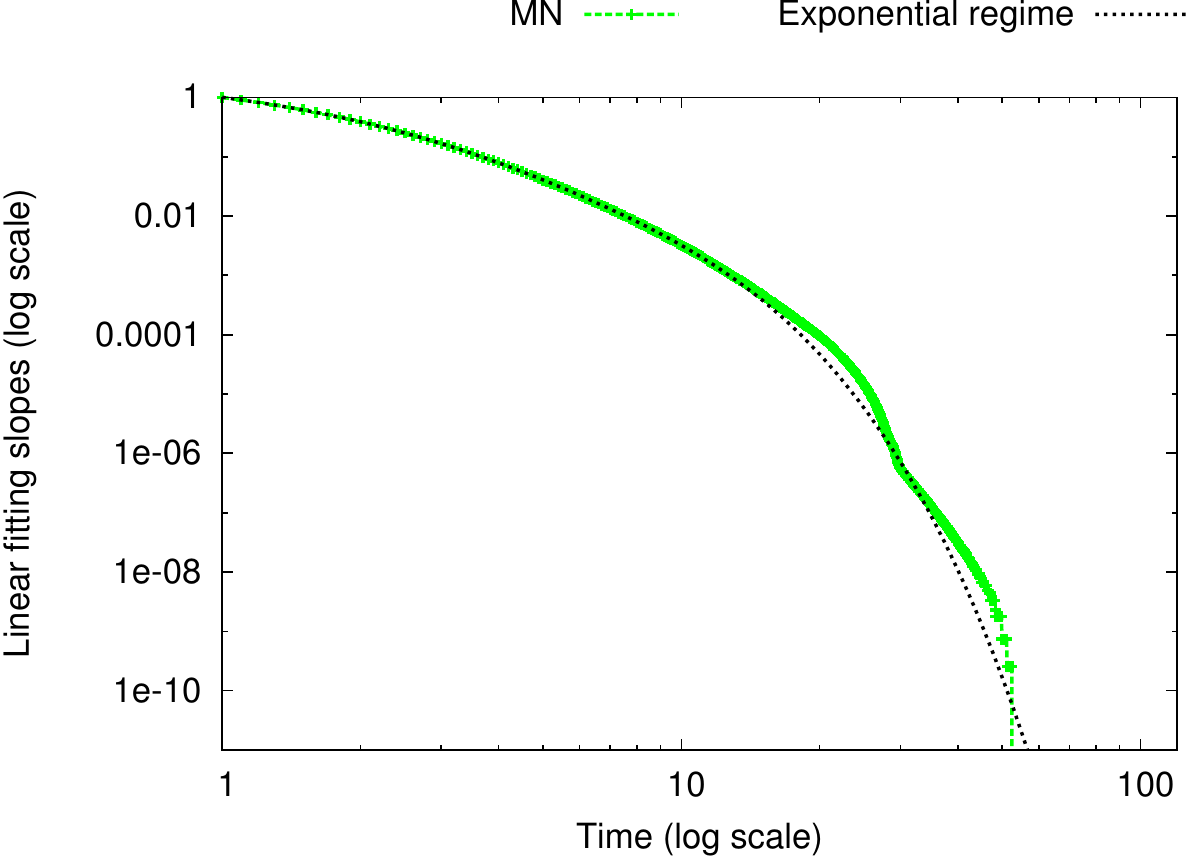}
\label{fig:scaling_HT_slopes_time__M}}
~
\subfigure[Linear fitting slopes of HT over time (GEN only).]{
\includegraphics[viewport=0 0 340 247,scale=0.6,keepaspectratio=true]{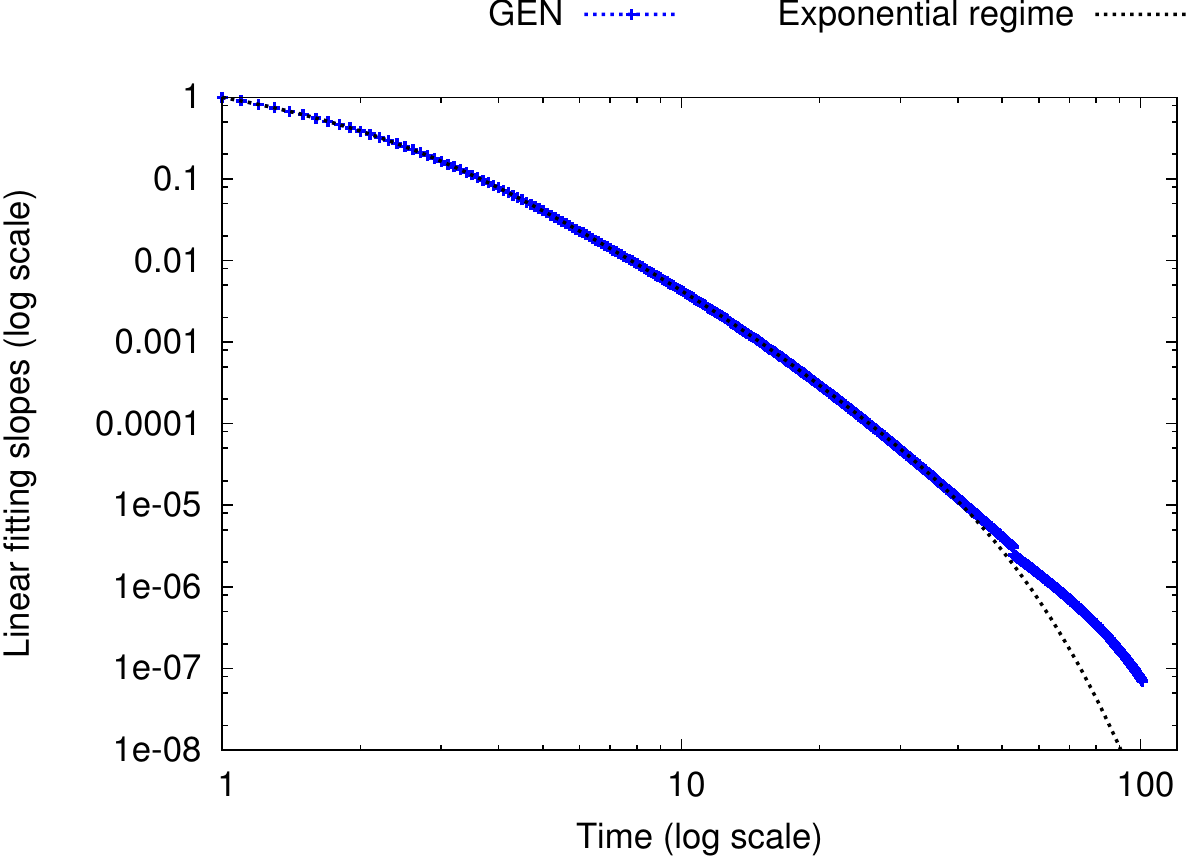}
\label{fig:scaling_HT_slopes_time__G}}

\caption{Scaling of HT linear best fitting slopes over time (time is sampled in 1000 equally-spaced points between 0 and 100).}
\label{fig:scaling_HT_slopes}
\end{figure*}

\begin{figure*}[ht!]
\centering

\subfigure[Scaling of HC for $t=1$.]{
\includegraphics[viewport=0 0 351 245,scale=0.62,keepaspectratio=true]{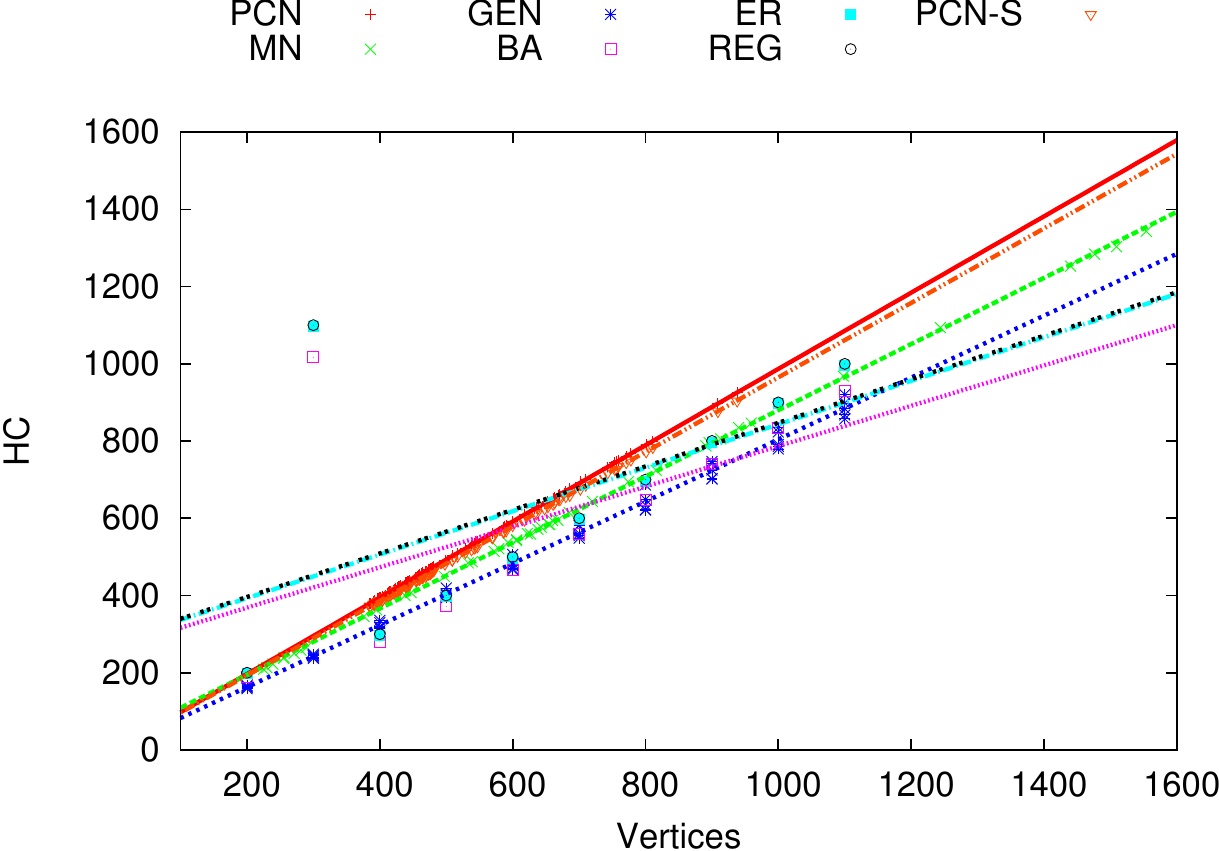}
\label{fig:scaling_HC_t1}}
~
\subfigure[Scaling of HC for $t=5$.]{
\includegraphics[viewport=0 0 351 245,scale=0.62,keepaspectratio=true]{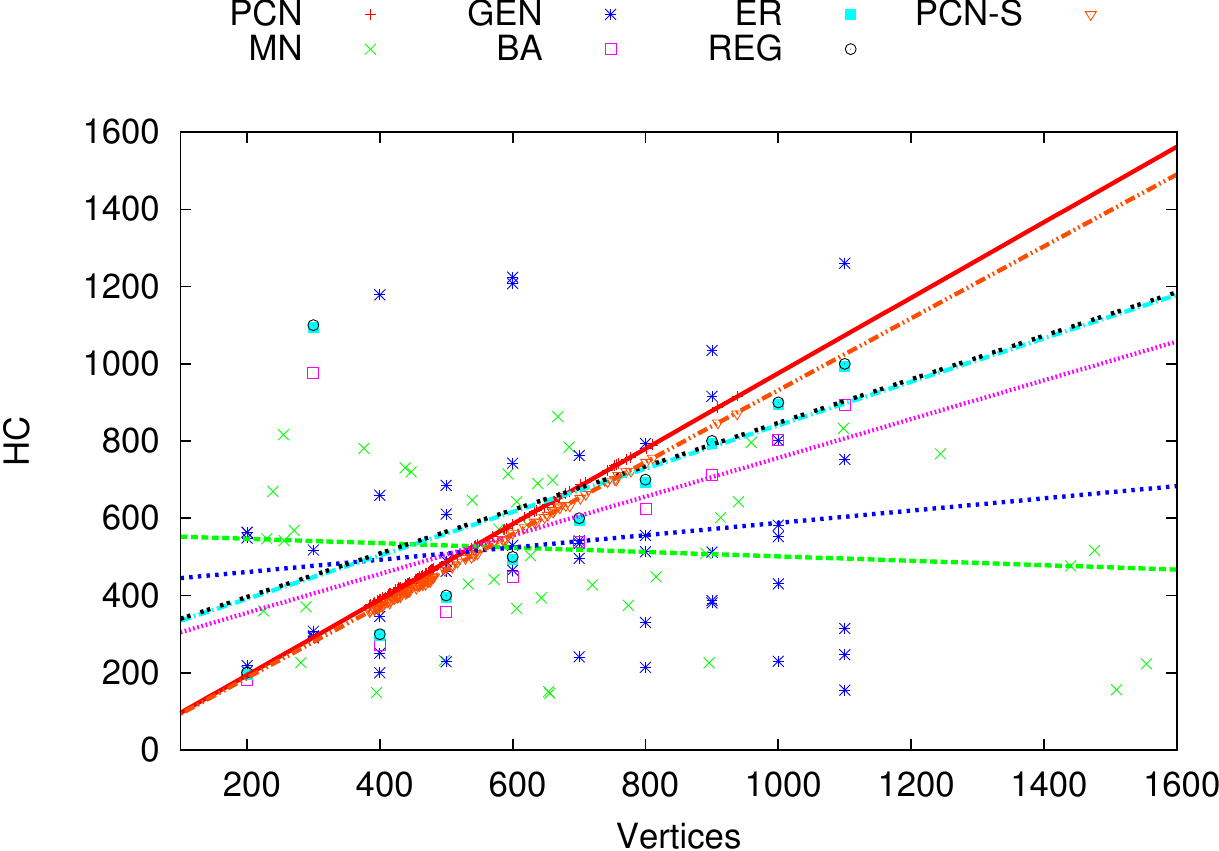}
\label{fig:scaling_HC_t5}}

\subfigure[Scaling of HC for $t=9$.]{
\includegraphics[viewport=0 0 351 245,scale=0.62,keepaspectratio=true]{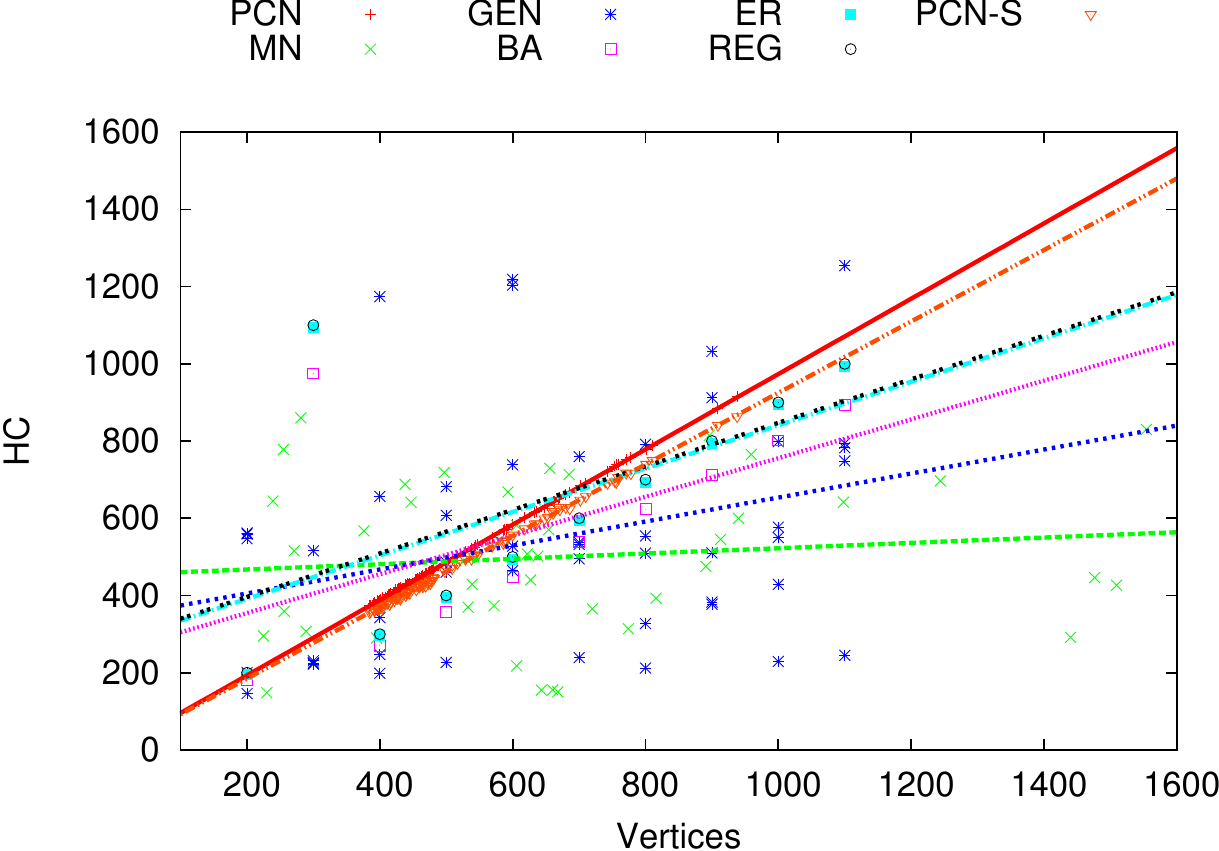}
\label{fig:scaling_HC_t9}}
~
\subfigure[Scaling of HCI (first coefficient).]{
\includegraphics[viewport=0 0 351 245,scale=0.62,keepaspectratio=true]{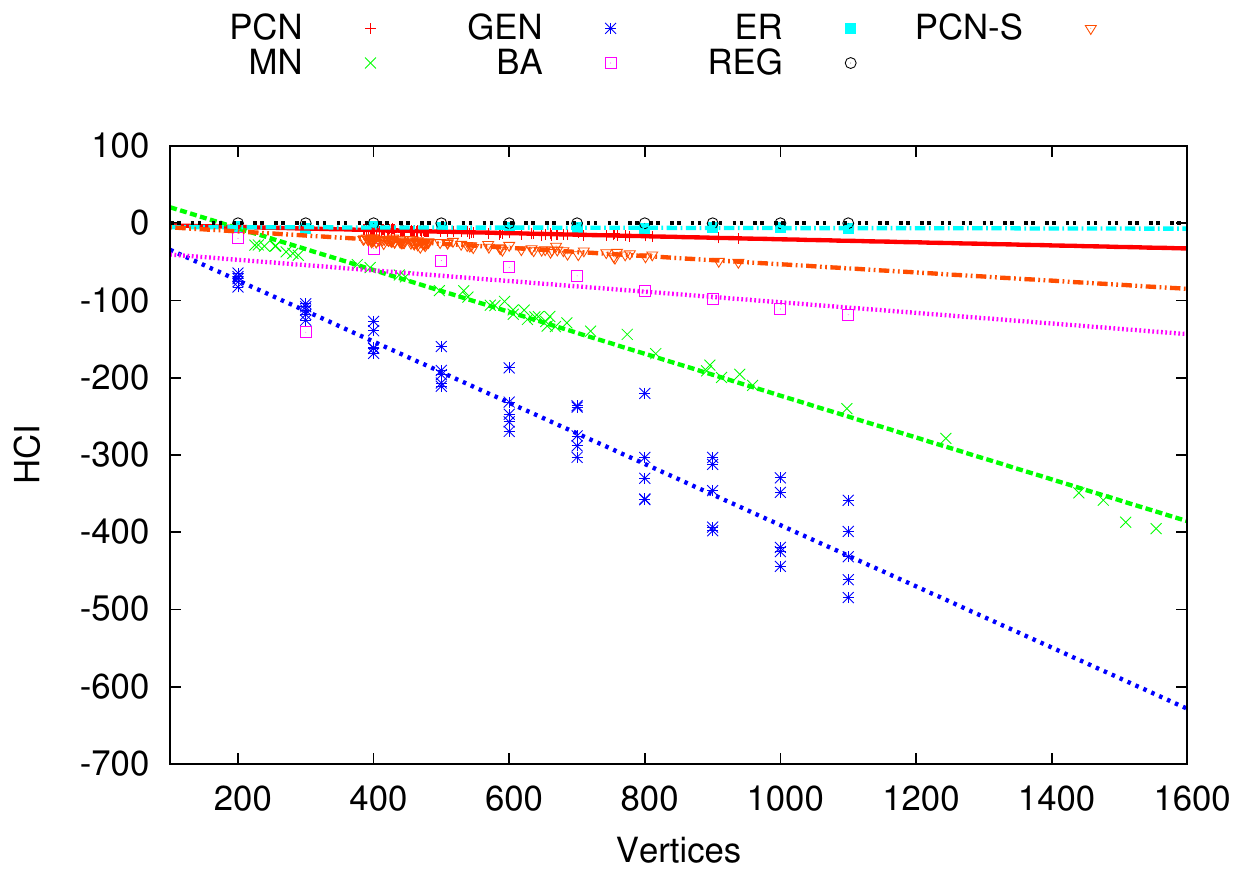}
\label{fig:scaling_HCI}}

\caption{Scaling of HC and HCI over network size.}
\label{fig:scaling_HC}
\end{figure*}

Figs. \ref{fig:scaling_HT}, \ref{fig:scaling_HT_slopes}, and \ref{fig:scaling_HC} show the scaling of all considered HK invariants.
Initially we consider only three relevant time instants for HT, i.e., $t=1, 5, 9$, which are depicted, respectively, in Figs. \ref{fig:scaling_HT_t1}, \ref{fig:scaling_HT_t5}, and \ref{fig:scaling_HT_t9}.
It is possible to note that, as expected, at $t=1$ all networks show a similar increasing linear trend w.r.t. the network size.
As the time instant increases, instead, PCN show a positive slope at least one order of magnitude greater than the others.
At first, this fact might be attributed solely to the intrinsic high modularity characterizing the protein structures.
To this end, in Fig. \ref{fig:correlations} we globally correlated MOD with HT over time -- the time $t$ here has a fine-grained sampling going from 0.1 to 100 with an increment step of 0.1. In the same plot, we show also the partial correlation obtained when considering the number of vertices as the control variable (indicated as ``MOD--HT(t) / V'' in the figure).
The linear correlation trend shows that initially the two quantities are fairly anti-correlated, while they soon become very correlated, reaching the maximum correlation ($\simeq 0.88$) around the time instant $t=10$. Successively, the correlation decreases with a smooth trend.
The partial correlation demonstrates that the initial negative correlation is due to the network size effect; correlations are positive when the size is removed.
This variability in the correlation points out the fact that, although the information provided by HT is fairly consistent with the one provided by MOD, they are by no means equivalent. Notably, HT offers a richer type of information that we will further exploit in the following in terms of diffusion.
In addition, this particular trend justifies the selection of the first ten time (integer-valued) instants for the calculations of PCA-HT, PCA-HC, and PCA-HCI performed in Sec. \ref{sec:anal_hk}.

Now we elaborate over the diffusion dynamics of the considered network ensembles by analyzing the log-log plots shown in Fig. \ref{fig:scaling_HT_slopes}.
Let us denote the time-dependent linear best-fitting (as those in Figs. \ref{fig:scaling_HT_t1}, \ref{fig:scaling_HT_t5}, and \ref{fig:scaling_HT_t9}) for the HT over the networks of a specific class, $\mathcal{C}$, as a function of the network size $n$, with the following expression:
\begin{equation}
\label{eq:linearfit_ht}
\text{HT}_\mathcal{C}(n; t) \simeq \alpha_{\mathcal{C}}(t)\cdot n,
\end{equation}
where $n$ is the number of vertices and $\alpha_{\mathcal{C}}(t)\geq 0$ is the time-dependent slope.
As discussed in Appendix \ref{sec:heat_kernel_spectrum}, by fitting linearly the HT we implicitly hypothesize the possibility to consistently describe each class of networks with an ensemble, $\mathcal{C}$, characterized by a unique probability density function of the normalized Laplacian eigenvalues (see Ref. \cite{mitrovic2009spectral} for a related theoretical study on a similar topic). This assumption is also justified by the results of PCA-HT reported in Fig. \ref{fig:PCA_HT}, which show good agreement among networks of the same class.
As a consequence, the linear best-fitting (\ref{eq:linearfit_ht}) allows to consider a statistic over an entire homogeneous class of networks, instead of focusing on each isolated network dynamics separately.
It is straightforward to realize that $\text{HT}_\mathcal{C}(n; t) = n$ for $t=0$, i.e., $\alpha_{\mathcal{C}}(0)=1$.
As $t$ grows, $\alpha_{\mathcal{C}}(t)$ becomes always smaller, with a rate that is related to the characteristic HT decay of the ensemble.
In Fig. \ref{fig:scaling_HT_slopes_time} we show the linear best-fitting slopes of HT, $\alpha_{\mathcal{C}}(t)$, as a function of time -- note that $t$ always varies from 0 to 100 with an increment step of 0.1.
While one expects to observe trends consistent with an exponential decay (see definition of HT in Eq. \ref{eq:heat_trace}), it is possible to recognize a different asymptotic trend for the PCN ensemble. For the sake of a better visualization, in Figs. \ref{fig:scaling_HT_slopes_time__P}, \ref{fig:scaling_HT_slopes_time__M}, and \ref{fig:scaling_HT_slopes_time__G} we report the same plot but isolating, respectively, PCN, MN, and GEN; other networks are omitted for brevity.
Fig. \ref{fig:scaling_HT_slopes_time__P} depicts what we might consider a change of functional form for the PCN trend at some point in time (i.e., starting around $t\simeq 5$).
This change of regime in the diffusion lasts few time instants, then the trend switches from exponential to power law like. This is not noted for the other networks that, instead, remain consistent with an exponential decay -- they are actually expressed as sums of exponentials.
In practice, from a certain $\widetilde{t}>5$, the diffusion in PCN seems to be consistent with a power law, $\alpha_{\mathcal{C}}(\widetilde{t}) \sim \widetilde{t}^{-\beta}$, where in our case the characteristic exponent is $\beta\simeq 1.1$.
It is worth mentioning that we achieved the very same results for PCN constructed without considering the lower filter at 4 \AA{}, that is, by considering all contacts within 8 \AA{} (data not shown).

Similar anomalies of functional form have been observed in the (cumulative) distribution of many experimental time series, especially in those related to financial markets \cite{kwapien2012physical}.
This phenomenon might happen when the functional form is consistent with one of the $q$-exponentials family, which originated in the field of non-extensive statistical mechanics \cite{tsallis2001nonextensive}.
In the case of PCN, this behavior is the signature of a crucial physical property of proteins, i.e., the energy flow.
Energy flow in proteins mimics the transport in a three-dimensional percolation cluster \cite{doi:10.1146/annurev.physchem.59.032607.093606}: energy flows readily between connected sites of the cluster and only slowly between non connected sites.
This experimentally validated double regime seems to be captured by the HT decay trend shown in Fig. \ref{fig:scaling_HT_slopes_time__P}. 
We stress that this result is elaborated from the herein exploited minimalistic PCN model, so confirming the relevance of graph-based representations in protein science.

Now let us consider the results for the HC (Figs. \ref{fig:scaling_HC_t1}, \ref{fig:scaling_HC_t5}, and \ref{fig:scaling_HC_t9}). Those three figures depict the scaling of the HC over the network size, considering the information of the entire HM. Notably, PCN and PCN-S are the only network types showing a consistent linear scaling with the size for all time instants. Other networks are not well-described by a linear fit as the time increases.
Finally, in Fig. \ref{fig:scaling_HCI} we show the scaling of the first HCI coefficient with the vertices (please note that for $m=1$, Eq. \ref{eq:hci} yields negative values). Of notable interest is the fact that PCN denote a nearly constant trend.
This means that, since the HCIs are time-independent features synthetically describing the HC information, PCN denote a similar characteristic in this respect, as in fact HC scaling in Fig. \ref{fig:scaling_HC} is consistently preserved over time.

\begin{figure*}[ht!]
\centering

\subfigure[HM for JW0058 at $t=1$.]{
\includegraphics[viewport=0 0 1043 634,scale=0.205,keepaspectratio=true]{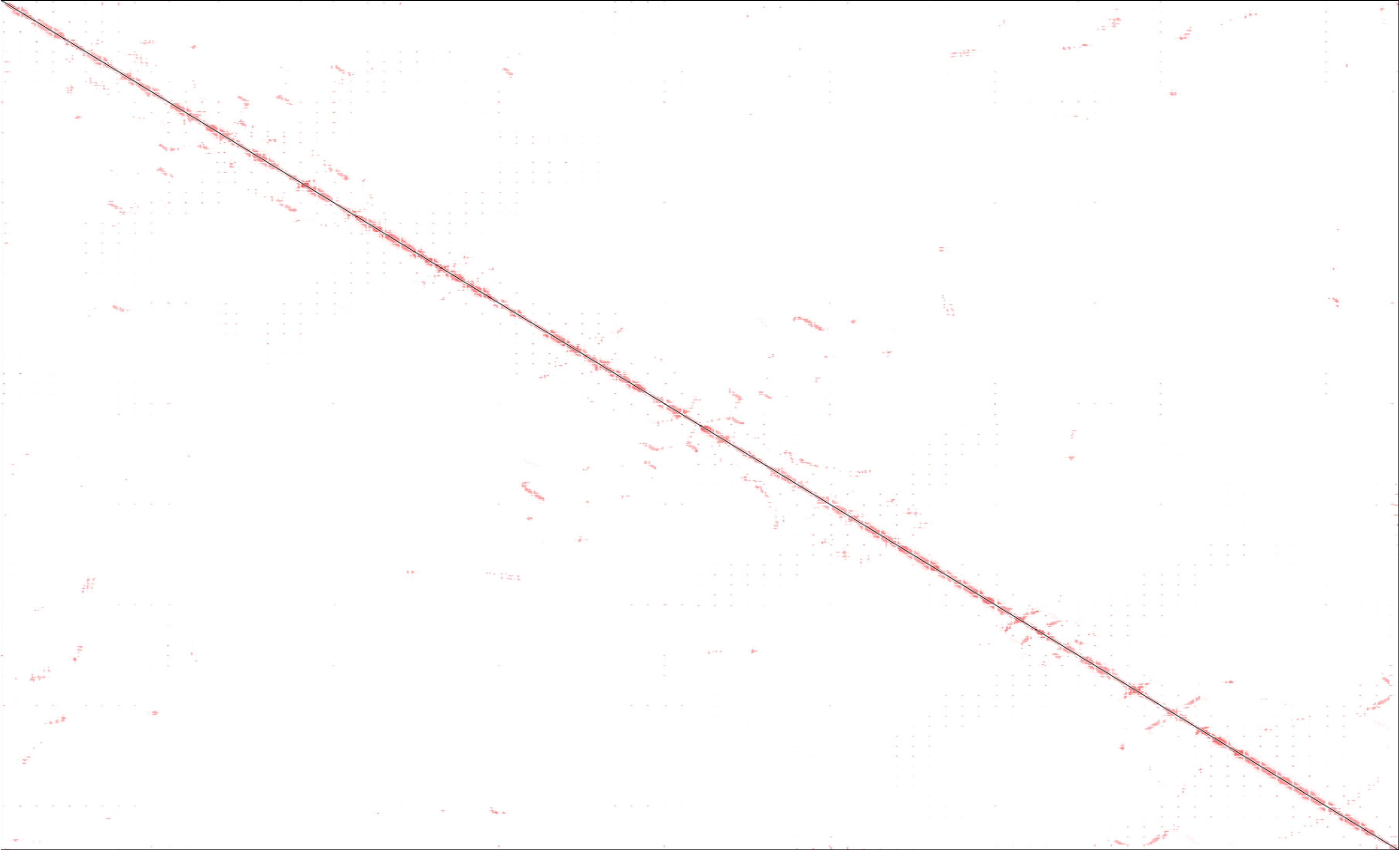}
\label{fig:HM_JW0058_t1}}
~
\subfigure[HM for JW0058-SYNTH at $t=1$.]{
\includegraphics[viewport=0 0 1037 615,scale=0.21,keepaspectratio=true]{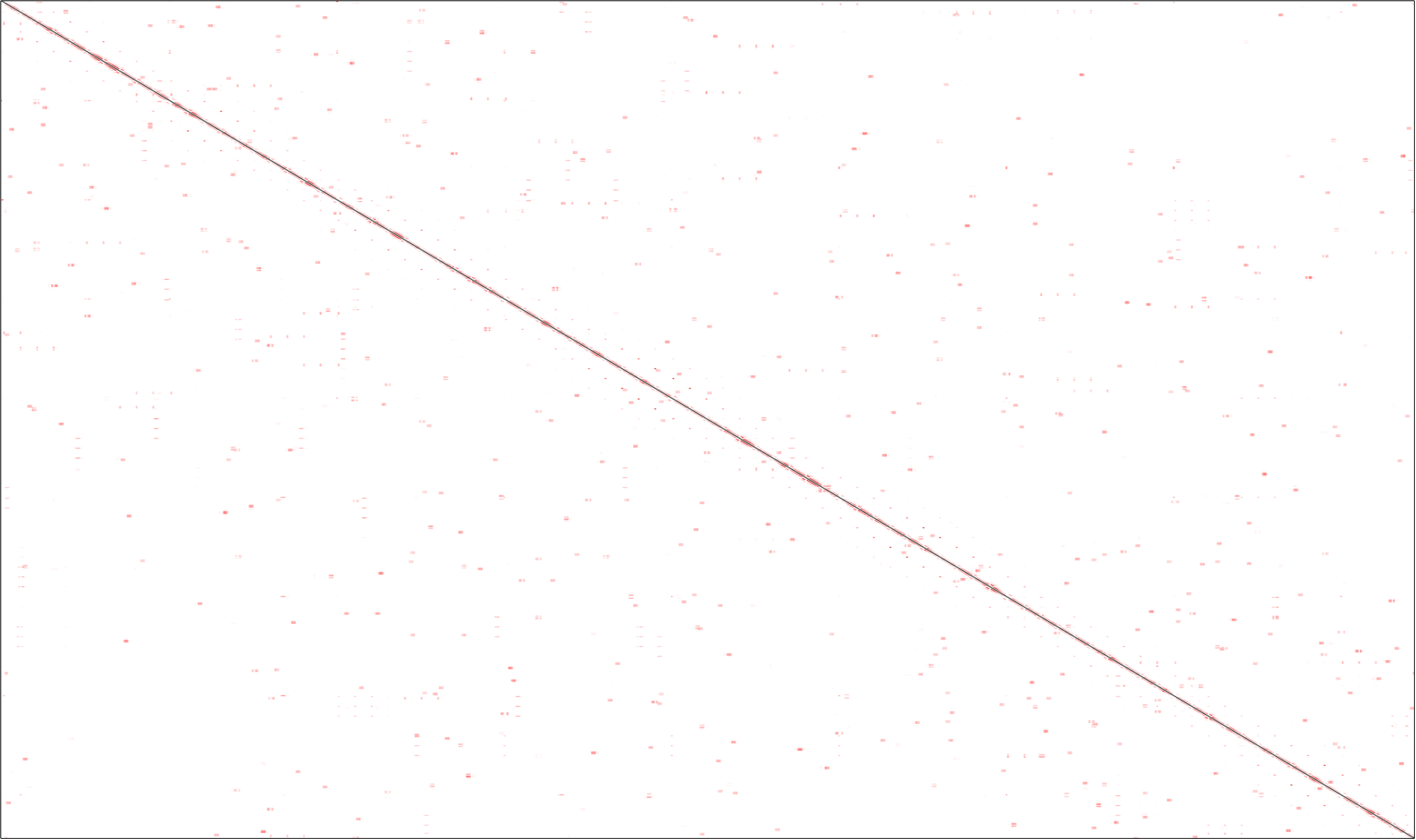}
\label{fig:HM_TY_t1}}

\subfigure[HM for JW0058 at $t=5$.]{
\includegraphics[viewport=0 0 1037 615,scale=0.21,keepaspectratio=true]{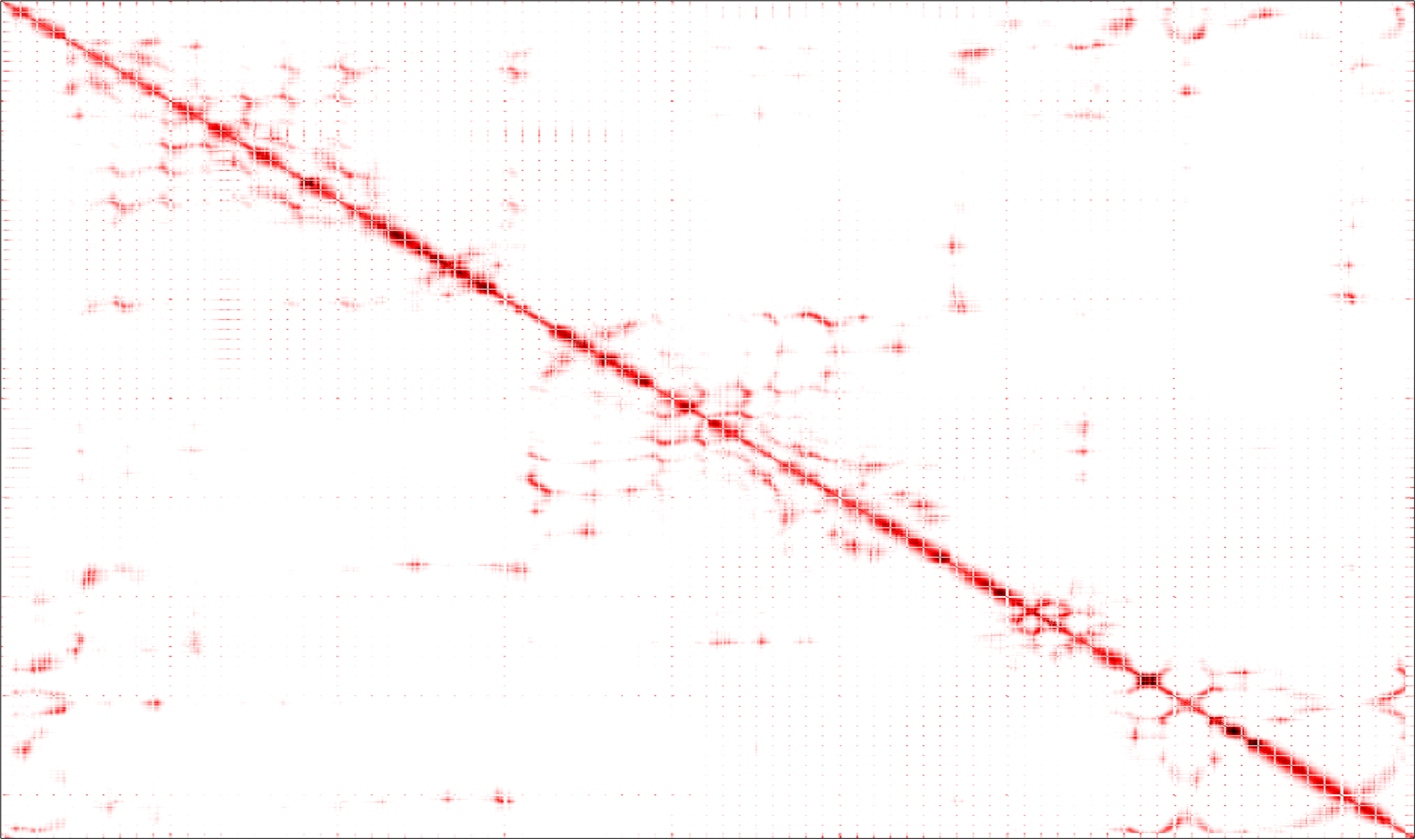}
\label{fig:HM_JW0058_t5}}
~
\subfigure[HM for JW0058-SYNTH at $t=5$.]{
\includegraphics[viewport=0 0 1043 634,scale=0.205,keepaspectratio=true]{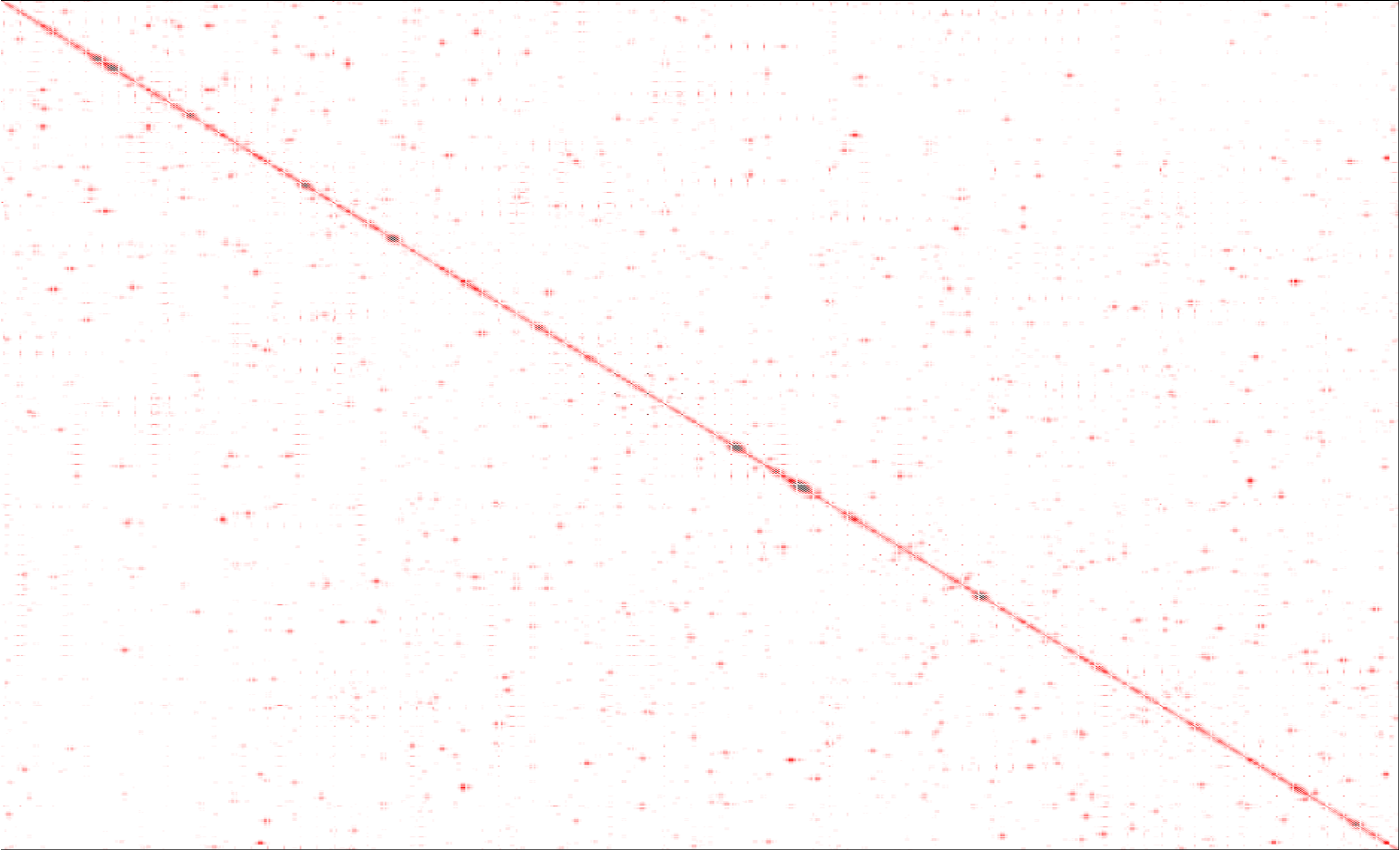}
\label{fig:HM_TY_t5}}

\subfigure[HM for JW0058 at $t=9$.]{
\includegraphics[viewport=0 0 1037 615,scale=0.21,keepaspectratio=true]{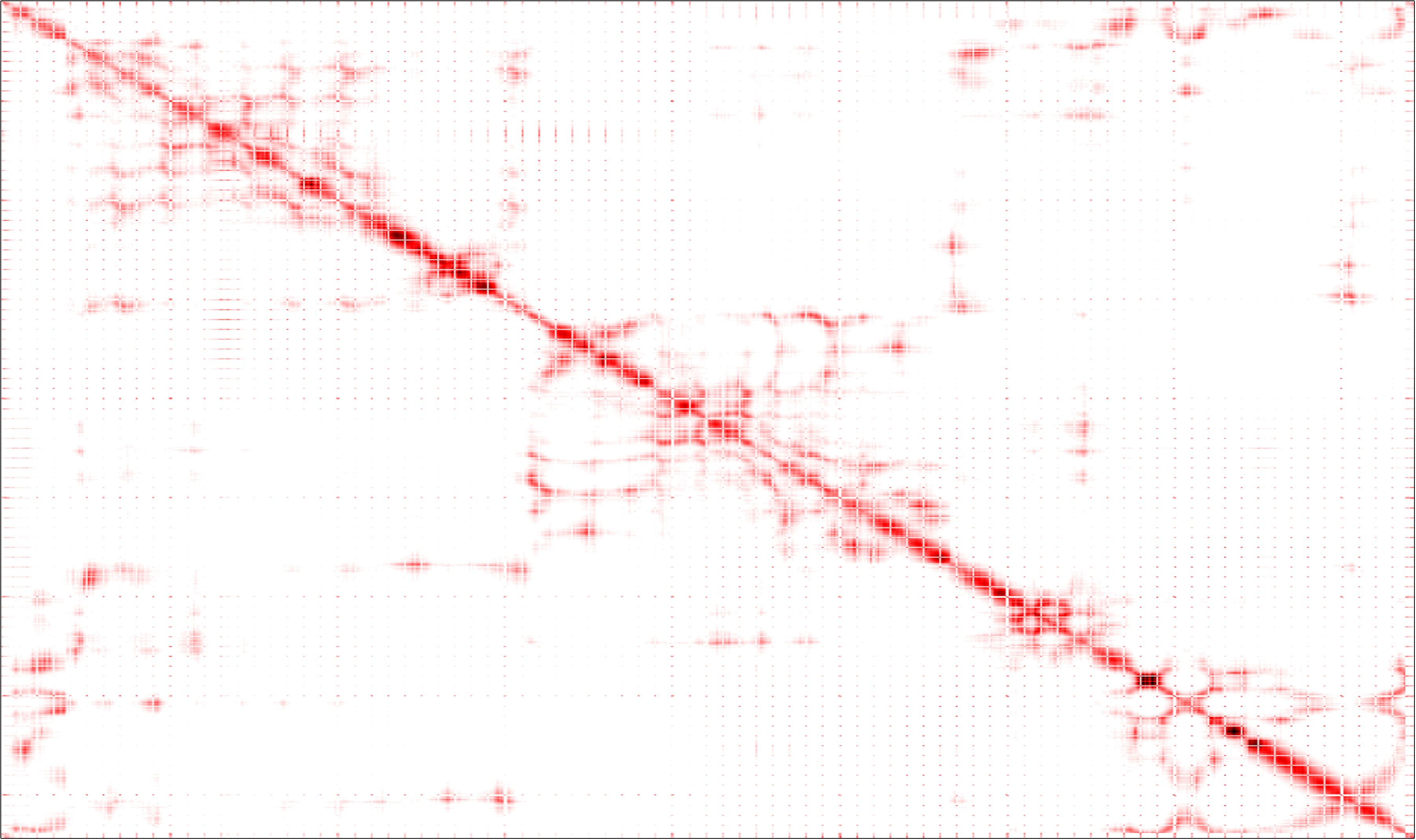}
\label{fig:HM_JW0058_t9}}
~
\subfigure[HM for JW0058-SYNTH at $t=9$.]{
\includegraphics[viewport=0 0 1037 615,scale=0.21,keepaspectratio=true]{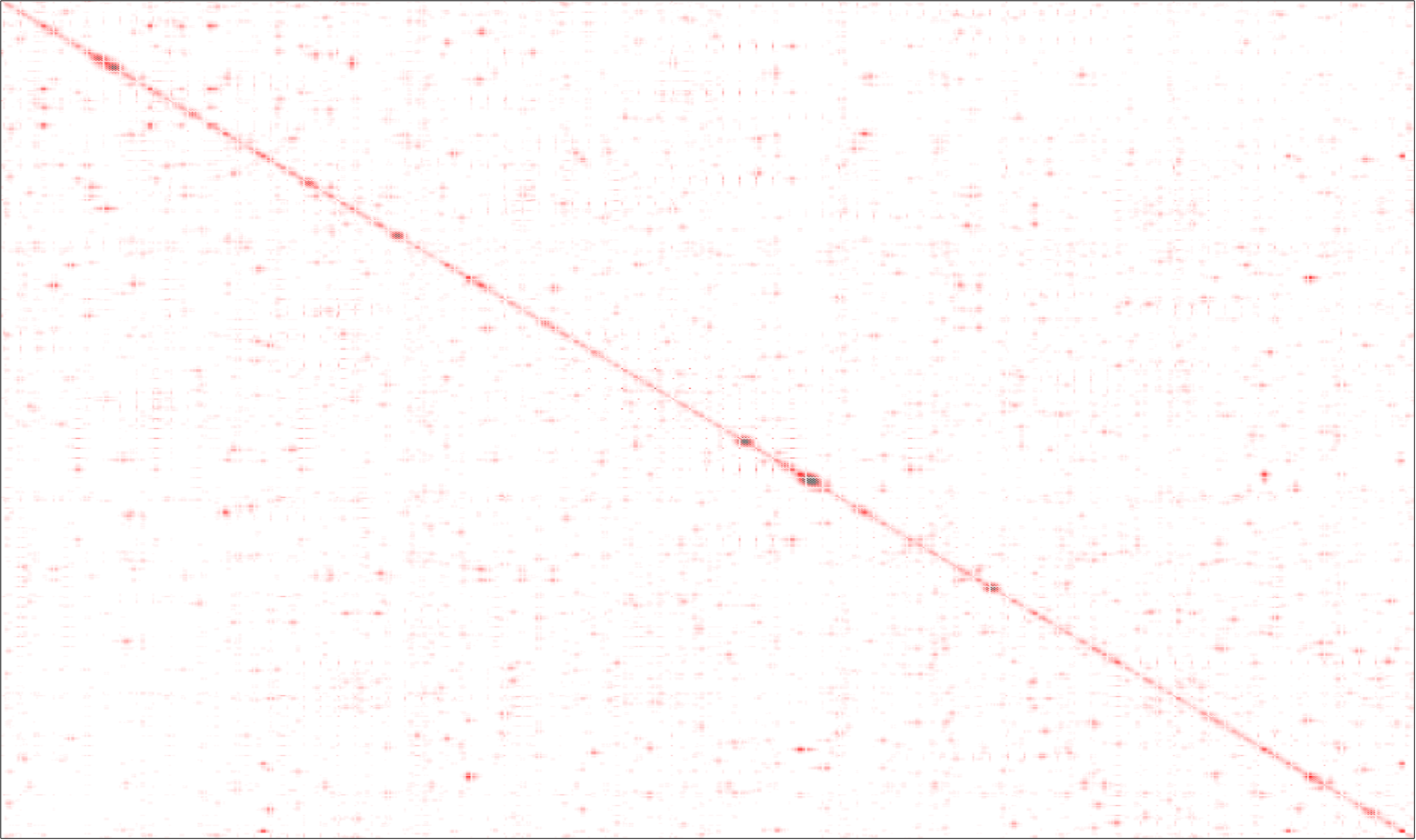}
\label{fig:HM_TY_t9}}

\subfigure[HM for JW0058 at $t=15$.]{
\includegraphics[viewport=0 0 1037 615,scale=0.21,keepaspectratio=true]{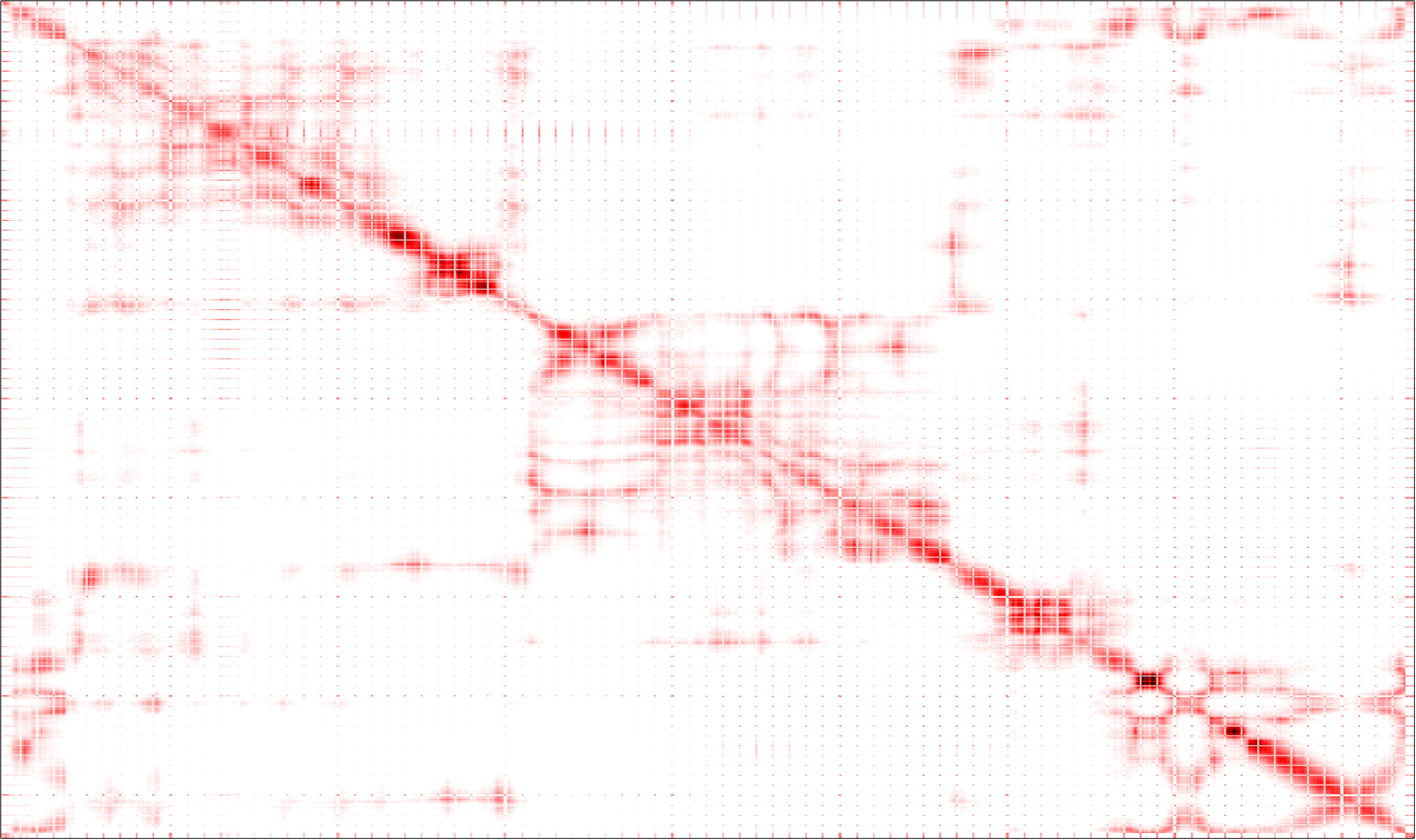}
\label{fig:HM_JW0058_t15}}
~
\subfigure[HM for JW0058-SYNTH at $t=15$.]{
\includegraphics[viewport=0 0 1037 615,scale=0.21,keepaspectratio=true]{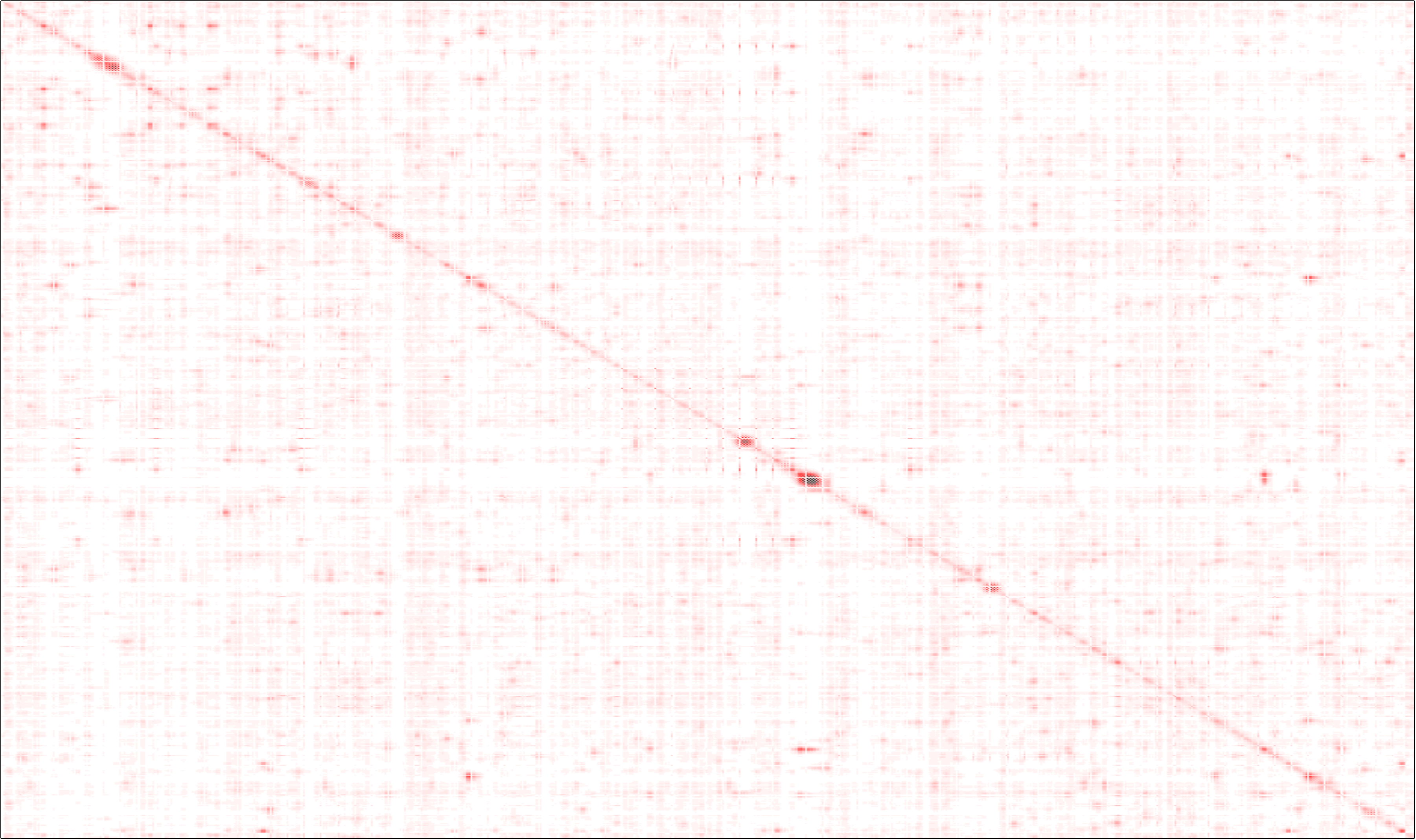}
\label{fig:HM_TY_t15}}

\caption{HM diffusion pattern over time for protein JW0058 and its synthetic counterpart.}
\label{fig:HM_time}
\end{figure*}

In Fig. \ref{fig:HM_time} we offer a visual representation of the heat diffusion pattern over time that is observable through the entire HM.
We considered two exemplar networks of exactly the same size: the ``JW0058'' protein and the synthetic counterpart belonging to PCN-S that we denote here as ``JW0058-SYNTH''.
As discussed before, PCN are characterized by a highly modular and fractal structure, while the considered synthetic counterpart exhibits a typical small world topology.
Accordingly, by comparing the diffusion occurring on the two networks over time, it is possible to recognize significantly different patterns that were not noted in the scalings of Fig. \ref{fig:scaling_HC}.
Of course, initially ($t=1$) the heat is mostly concentrated in the vertices, which results in a very intense trace.
As the time increases, the diffusion pattern for the real protein is more evident and also persistent.
This is in agreement with recent laboratory experiments \cite{doi:10.1146/annurev.physchem.59.032607.093606,lervik2010heat}, which demonstrated that diffusion in proteins proceeds slower than normal diffusion.
Conversely, the diffusion for JW0058-SYNTH is in general faster since in fact the trace vanishes quickly.
In graph-theoretical terms, this means that the spectral gap of PCN dominates Eq. \ref{eq:heat_matrix} as $t$ becomes large.
This result suggests us that, in future research studies, it could be interesting to devote focused attention to the properties of the spectral density of the ensembles.
Analogue results have been obtained by considering the other network types; we do not show them for the sake of brevity.

\clearpage
\section{Conclusions}
\label{sec:conclusions}

In this paper we have investigated the structure of three types of complex networks: protein contact networks, metabolic networks, and gene regulatory networks, together with simulated archetypal models acting as probes.
We biased the study on protein contact networks, highlighting their peculiar structure with respect to the other networks.
Our analysis focused on ensemble statistics, that is, we analyzed the features elaborated by considering several instances of varying size of such networks.
We considered two main network characterizations: the first one based on classical topological descriptors, while the second one exploited several invariants extracted from the discrete heat kernel.
We found strong statistical agreement among those two representations, which allowed for a consistent interpretation of the results in terms of principal component analysis.
Our major result was the demonstration of a double regime characterizing a (simulated) diffusion process in the considered protein contact networks.
As shown by laboratory experiments, both energy flow and vibration dynamics in proteins exhibit subdiffusive properties, i.e., slower-than-normal diffusion \cite{doi:10.1146/annurev.physchem.59.032607.093606}.
The notable difference in the diffusion pattern between real proteins and the herein considered simulated polymers (whose contact networks have the same local structure of the corresponding real proteins), points to a peculiar mesoscopic organization of proteins going beyond the pure backbone folding.
The observed correlations between MOD and HT indicates this principle in the presence of well-characterized domains.
The novelty of our results is that we were able to demonstrate such a well-known property of proteins by exploiting graph-based representations and computational tools only.
The fact that the observed properties emerged with no explicit reference to chemico-physical characterization of proteins, relying hence on pure topological properties, suggests the existence of general universal mesoscopic principles fulfilling the hopes expressed by Laughlin \textit{et al.} \cite{laughlin2000middle}.

\appendix
\section{Heat Kernel and the Related Invariants}
\label{sec:heat_kernel}

Let $G=(\mathcal{V}, \mathcal{E})$ be a graph (network) with $n=|\mathcal{V}|$ vertices and $m=|\mathcal{E}|$ edges.
Let $\mathbf{A}^{n\times n}$ be the adjacency matrix defined as $A_{ij}=1$ if there is an edge between vertices $v_i,v_j\in\mathcal{V}$; $A_{ij}=0$ otherwise. Let us define the degree of a vertex $v_i$ as $\mathrm{deg}(v_i)=\sum_{j=1}^{n} A_{ij}$. In addition, let us define $\mathbf{D}$ as a diagonal matrix of degree: $D_{ii}=\mathrm{deg}(v_i)$.
Let $\mathbf{L}=\mathbf{D}-\mathbf{A}$ be the Laplacian matrix of $G$.
The normalized Laplacian matrix is given by $\hat{\mathbf{L}}=\mathbf{D}^{-1/2}\mathbf{L}\mathbf{D}^{-1/2}$, and as a consequence $\hat{\mathbf{L}}$ is symmetric and positive semi-definite; therefore it has non-negative eigenvalues only.
Let us define the spectral decomposition of the Laplacian as $\hat{\mathbf{L}}=\Phi \Lambda \Phi^{T}$, where $\Lambda$ is the diagonal matrix containing the eigenvalues arranged as $0=\lambda_1\leq \lambda_2\leq ...\leq \lambda_n\leq 2$; $\Phi$ contains the corresponding (unitary) eigenvectors as columns.

The heat equation \cite{Xiao:2009:GCH:1563046.1563099} associated to the normalized Laplacian, $\hat{\mathbf{L}}$, is given by
\begin{equation}
\label{eq:heat_equation}
\frac{\partial \mathbf{H}_{t}}{\partial t} = -\hat{\mathbf{L}}\mathbf{H}_{t},
\end{equation}
where $\mathbf{H}_{t}$ is a doubly-stochastic $n\times n$ matrix, called heat matrix (HM), and $t$ is the time variable.
It is well-known that the solution to (\ref{eq:heat_equation}) is
\begin{equation}
\mathbf{H}_{t} = \exp(-\hat{\mathbf{L}}t),
\end{equation}
which can be solved by exponentiating the spectrum of $\hat{\mathbf{L}}$:
\begin{equation}
\mathbf{H}_{t} = \Phi \exp(-\Lambda t) \Phi^{T} = \sum_{i=1}^{n} \exp(-\lambda_i t)\phi_i\phi_i^{T}.
\end{equation}

Eq. \ref{eq:heat_equation} describes the diffusion of heat/information across the graph over time. In fact,
\begin{equation}
\label{eq:heat_matrix}
\mathbf{H}_{t}(v, u) = \sum_{i=1}^{n} \exp(- \lambda_i t)\phi_i(v)\phi_i(u),
\end{equation}
where $\phi_i(v)$ is the value related to the vertex $v$ in the \textit{i}th eigenvector.
It is important to note that $\mathbf{H}_{t} \simeq \mathbf{I} - \hat{\mathbf{L}}t$ when $t\rightarrow 0$; conversely, when $t$ is large $\mathbf{H}_{t}\simeq \exp(-\lambda_2 t)\phi_2\phi_{2}^{T}$, where $\phi_2$ is the normalized Fiedler vector.
This means that the large-time behavior of the diffusion depends on the global structure of the graph (e.g., its global architectural organization), while its short-time characteristics are determined by the local structure.

The heat trace (HT) of $\mathbf{H}_{t}$ is given by
\begin{equation}
\label{eq:heat_trace}
\mathrm{HT}(t) = \mathrm{Tr}(\mathbf{H}_{t}) = \sum_{i=1}^{n} \exp(-\lambda_i t),
\end{equation}
which takes into account only the eigenvalues of $\hat{\mathbf{L}}$.
The heat content (HC) of $\mathbf{H}_t$ is defined by considering also the eigenvectors of $\hat{\mathbf{L}}$:
\begin{align}
\label{eq:heat_content}
\mathrm{HC}(t) = \sum_{u\in\mathcal{V}} \sum_{u\in\mathcal{V}} \mathbf{H}_{t}(u, v) = \sum_{u\in\mathcal{V}} \sum_{u\in\mathcal{V}} \sum_{i=1}^{n} \exp(-\lambda_i t)\phi_i(v)\phi_i(u).
\end{align}

Eq. \ref{eq:heat_content} can be described in terms of power series expansion,
\begin{equation}
\label{eq:heat_coeff}
\mathrm{HC}(t) = \sum_{m=0}^{\infty} q_m t^m.
\end{equation}

By using the McLaurin series for the exponential function, we have
\begin{equation}
\exp(-\lambda_i t)=\sum_{m=0}^{\infty} \frac{(-\lambda_i)^{m} t^m}{m!},
\end{equation}
which substituted in Eq. \ref{eq:heat_content} gives:
\begin{align}
\label{eq:heat_content2}
\mathrm{HC}(t) = \sum_{u\in\mathcal{V}} \sum_{u\in\mathcal{V}} \sum_{i=1}^{n} \exp(-\lambda_i t)\phi_i(v)\phi_i(u) = \sum_{m=0}^{\infty}\sum_{u\in\mathcal{V}} \sum_{u\in\mathcal{V}} \sum_{i=1}^{n} \phi_i(v)\phi_i(u) \frac{(-\lambda_i)^{m} t^m}{m!}.
\end{align}

The $q_m$ coefficients in (\ref{eq:heat_coeff}) are graph invariants (called heat content invariants, HCI) that can be calculated in closed-form by using Eqs. \ref{eq:heat_coeff} and \ref{eq:heat_content2}:
\begin{equation}
\label{eq:hci}
q_m = \displaystyle\sum_{i=1}^{n} \left( \sum_{u\in\mathcal{V}} \phi_{i}(u) \right)^2 \frac{(-\lambda_i)^m}{m!}.
\end{equation}

\section{Ensemble Heat Trace}
\label{sec:heat_kernel_spectrum}

The HT (\ref{eq:heat_trace}) of a graph $G=(\mathcal{V}, \mathcal{E})$ with $n = |\mathcal{V}|$ can be expressed as a function of time
\begin{equation}
\label{eq:httv}
\text{HT}_G(t) = \sum_{i=1}^n \exp(- \lambda_i t ) = 1 + \sum_{i=2}^n \exp(- \lambda_i t ) .
\end{equation}
where $\lambda_i$ are the eigenvalues of the normalized Laplacian of $G$.
Let us define an ensemble of graphs $\mathcal{C}$, in which all graphs share a common characteristic spectral density. Such spectra can be synthetically described by considering the spectral density of the ensemble $\mathcal{C}$. Accordingly, we can consider the eigenvalues as i.i.d. random variables $\tilde{\lambda}_i$, assuming values according to the spectral density of the ensemble -- except for $\tilde{\lambda}_1$ that assumes deterministically the value 0.
The HT of a generic graph $G \in \mathcal{C}, n=|\mathcal{V}|,$ can be written as:
\begin{equation}
\label{eq:ht_graph}
\text{HT}_G (t; n) = 1 + \sum_{i=2}^n \exp(- \tilde{\lambda}_i t ) = 1 + \sum_{i=2}^n \exp(- \tilde{\lambda} t).
\end{equation}

The last step in Eq. \ref{eq:ht_graph} is carried out by considering that, since the $\tilde{\lambda}_i$ are assumed as i.i.d., their values can be actually expressed as $n$ realizations of a single random variable, $\tilde{\lambda}$.
For a fixed value of time $t$, we can define the \textit{ensemble} HT, $\text{HT}_\mathcal{C}(n; t)$, as the mean HT over all graphs of the ensemble $\mathcal{C}$ with varying size $n$, which is given by:
\begin{equation}
\text{HT}_\mathcal{C}(n; t) = \langle \text{HT}_G(t; n) \rangle_\mathcal{C} = 1 + \sum_{i=2}^n \langle \exp(- \tilde{\lambda} t ) \rangle_\mathcal{C} = 1 + (n-1) \langle \exp(- \tilde{\lambda} t ) \rangle_\mathcal{C}.
\end{equation}

Hence, $\text{HT}_\mathcal{C}(n; t)$ can be expressed as a linear function of the graph size
\begin{equation}
\text{HT}_\mathcal{C}(n; t) = 1 - \alpha_\mathcal{C}(t) + \alpha_\mathcal{C}(t) \cdot n \simeq \alpha_\mathcal{C}(t) \cdot n,
\end{equation}
where $\alpha_\mathcal{C}(t) = \langle \exp{(- \tilde{\lambda} t)} \rangle_\mathcal{C}\in[0, 1]$ is a time-dependent angular coefficient (slope) that is characteristic for the entire ensemble $\mathcal{C}$.

\bibliographystyle{abbrvnat}
\bibliography{/home/lorenzo/University/Research/Publications/Bibliography.bib}

\begin{thebibliography}{81}
\providecommand{\natexlab}[1]{#1}
\providecommand{\url}[1]{\texttt{#1}}
\expandafter\ifx\csname urlstyle\endcsname\relax
  \providecommand{\doi}[1]{doi: #1}\else
  \providecommand{\doi}{doi: \begingroup \urlstyle{rm}\Url}\fi

\bibitem[dre()]{dream5}
{DREAM5}.
\newblock URL \url{http://wiki.c2b2.columbia.edu/dream/index.php/D5c3}.

\bibitem[pdb()]{pdb}
{Protein Data Bank}.
\newblock URL \url{http://www.rcsb.org/pdb/home/home.do}.

\bibitem[Anand and Bianconi(2009)]{PhysRevE.80.045102}
K.~Anand and G.~Bianconi.
\newblock Entropy measures for networks: Toward an information theory of
  complex topologies.
\newblock \emph{Physical Review E}, 80:\penalty0 045102, Oct 2009.
\newblock \doi{10.1103/PhysRevE.80.045102}.

\bibitem[Bagler and Sinha(2005)]{bagler2005network}
G.~Bagler and S.~Sinha.
\newblock Network properties of protein structures.
\newblock \emph{Physica A: Statistical Mechanics and its Applications},
  346\penalty0 (1):\penalty0 27--33, 2005.

\bibitem[Banavar and Maritan(2007)]{banavar2007physics}
J.~R. Banavar and A.~Maritan.
\newblock Physics of proteins.
\newblock \emph{Annual Review of Biophysics and Biomolecular Structure},
  36:\penalty0 261--280, 2007.

\bibitem[Banerji and Ghosh(2011)]{banerji2011fractal}
A.~Banerji and I.~Ghosh.
\newblock Fractal symmetry of protein interior: what have we learned?
\newblock \emph{Cellular and Molecular Life Sciences}, 68\penalty0
  (16):\penalty0 2711--2737, 2011.

\bibitem[Barab{\'a}si and Albert(1999)]{barabasi1999emergence}
A.-L. Barab{\'a}si and R.~Albert.
\newblock Emergence of scaling in random networks.
\newblock \emph{Science}, 286\penalty0 (5439):\penalty0 509--512, 1999.

\bibitem[Barabasi and Oltvai(2004)]{barabasi2004network}
A.-L. Barabasi and Z.~N. Oltvai.
\newblock Network biology: understanding the cell's functional organization.
\newblock \emph{Nature Reviews Genetics}, 5\penalty0 (2):\penalty0 101--113,
  2004.

\bibitem[Barth{\'e}lemy(2011)]{barthelemy2011spatial}
M.~Barth{\'e}lemy.
\newblock Spatial networks.
\newblock \emph{Physics Reports}, 499\penalty0 (1):\penalty0 1--101, 2011.

\bibitem[Bartoli et~al.(2007)Bartoli, Fariselli, and
  Casadio]{bartoli2007effect}
L.~Bartoli, P.~Fariselli, and R.~Casadio.
\newblock The effect of backbone on the small-world properties of protein
  contact maps.
\newblock \emph{Physical Biology}, 4\penalty0 (4):\penalty0 L1, 2007.

\bibitem[Bashan et~al.(2012)Bashan, Bartsch, Kantelhardt, Havlin, and
  Ivanov]{bashan2012network}
A.~Bashan, R.~P. Bartsch, J.~W. Kantelhardt, S.~Havlin, and P.~C. Ivanov.
\newblock Network physiology reveals relations between network topology and
  physiological function.
\newblock \emph{Nature communications}, 3:\penalty0 702, 2012.

\bibitem[Ben-Avraham and Havlin(2000)]{ben2000diffusion}
D.~Ben-Avraham and S.~Havlin.
\newblock \emph{{Diffusion and reactions in fractals and disordered systems}}.
\newblock Cambridge University Press, 2000.

\bibitem[Blondel et~al.(2008)Blondel, Guillaume, Lambiotte, and
  Lefebvre]{blondel2008fast}
V.~D. Blondel, J.-L. Guillaume, R.~Lambiotte, and E.~Lefebvre.
\newblock Fast unfolding of communities in large networks.
\newblock \emph{Journal of Statistical Mechanics: Theory and Experiment},
  2008\penalty0 (10):\penalty0 P10008, 2008.

\bibitem[Boccaletti et~al.(2006)Boccaletti, Latora, Moreno, Chavez, and
  Hwang]{boccaletti+latora+moreno+chavez+hwang2006}
S.~Boccaletti, V.~Latora, Y.~Moreno, M.~Chavez, and D.~Hwang.
\newblock Complex networks: Structure and dynamics.
\newblock \emph{Physics Reports}, 424\penalty0 (4-5):\penalty0 175--308, Feb
  2006.
\newblock ISSN 03701573.
\newblock \doi{10.1016/j.physrep.2005.10.009}.

\bibitem[B{\"o}de et~al.(2007)B{\"o}de, Kov{\'a}cs, Szalay, Palotai,
  Korcsm{\'a}ros, and Csermely]{bode2007network}
C.~B{\"o}de, I.~A. Kov{\'a}cs, M.~S. Szalay, R.~Palotai, T.~Korcsm{\'a}ros, and
  P.~Csermely.
\newblock Network analysis of protein dynamics.
\newblock \emph{Febs Letters}, 581\penalty0 (15):\penalty0 2776--2782, 2007.

\bibitem[Bullmore and Sporns(2012)]{bullmore2012economy}
E.~T. Bullmore and O.~Sporns.
\newblock The economy of brain network organization.
\newblock \emph{Nature Reviews Neuroscience}, 13\penalty0 (5):\penalty0
  336--349, 2012.
\newblock \doi{10.1038/nrn3214}.

\bibitem[Costa et~al.(2007)Costa, Rodrigues, Travieso, and {Villas
  Boas}]{costa2007characterization}
L.~d.~F. Costa, F.~A. Rodrigues, G.~Travieso, and P.~R. {Villas Boas}.
\newblock {Characterization of complex networks: A survey of measurements}.
\newblock \emph{Advances in Physics}, 56\penalty0 (1):\penalty0 167--242, 2007.

\bibitem[Csermely et~al.(2012)Csermely, {Singh Sandhu}, Hazai, Hoksza, {J M
  Kiss}, Miozzo, Veres, Piazza, and Nussinov]{csermely2012disordered}
P.~Csermely, K.~{Singh Sandhu}, E.~Hazai, Z.~Hoksza, H.~{J M Kiss}, F.~Miozzo,
  D.~V. Veres, F.~Piazza, and R.~Nussinov.
\newblock Disordered proteins and network disorder in network descriptions of
  protein structure, dynamics and function: hypotheses and a comprehensive
  review.
\newblock \emph{Current Protein and Peptide Science}, 13\penalty0 (1):\penalty0
  19--33, 2012.

\bibitem[Csermely et~al.(2013{\natexlab{a}})Csermely, Korcsm{\'a}ros, Kiss,
  London, and Nussinov]{csermely2013structure}
P.~Csermely, T.~Korcsm{\'a}ros, H.~J.~M. Kiss, G.~London, and R.~Nussinov.
\newblock Structure and dynamics of molecular networks: A novel paradigm of
  drug discovery: A comprehensive review.
\newblock \emph{Pharmacology \& therapeutics}, 138\penalty0 (3):\penalty0
  333--408, 2013{\natexlab{a}}.

\bibitem[Csermely et~al.(2013{\natexlab{b}})Csermely, London, Wu, and
  Uzzi]{csermely2013structuredynamics}
P.~Csermely, A.~London, L.-Y. Wu, and B.~Uzzi.
\newblock Structure and dynamics of core/periphery networks.
\newblock \emph{Journal of Complex Networks}, 1\penalty0 (2):\penalty0 93--123,
  2013{\natexlab{b}}.

\bibitem[Dehmer and Mowshowitz(2011)]{Dehmer201157}
M.~Dehmer and A.~Mowshowitz.
\newblock {A history of graph entropy measures}.
\newblock \emph{Information Sciences}, 181\penalty0 (1):\penalty0 57--78, 2011.
\newblock ISSN 0020-0255.
\newblock \doi{10.1016/j.ins.2010.08.041}.

\bibitem[Dehmer et~al.(2013)Dehmer, Mueller, and
  Emmert-Streib]{dehmer2013quantitative}
M.~Dehmer, L.~A.~J. Mueller, and F.~Emmert-Streib.
\newblock Quantitative network measures as biomarkers for classifying prostate
  cancer disease states: a systems approach to diagnostic biomarkers.
\newblock \emph{PloS one}, 8\penalty0 (11):\penalty0 e77602, 2013.

\bibitem[Delvenne et~al.(2010)Delvenne, Yaliraki, and
  Barahona]{delvenne2010stability}
J.-C. Delvenne, S.~N. Yaliraki, and M.~Barahona.
\newblock Stability of graph communities across time scales.
\newblock \emph{Proceedings of the National Academy of Sciences}, 107\penalty0
  (29):\penalty0 12755--12760, 2010.
\newblock \doi{10.1073/pnas.0903215107}.

\bibitem[{Di Paola} et~al.(2012{\natexlab{a}}){Di Paola}, {De Ruvo}, Paci,
  Santoni, and Giuliani]{doi:10.1021/cr3002356}
L.~{Di Paola}, M.~{De Ruvo}, P.~Paci, D.~Santoni, and A.~Giuliani.
\newblock Protein contact networks: an emerging paradigm in chemistry.
\newblock \emph{Chemical Reviews}, 113\penalty0 (3):\penalty0 1598--1613,
  2012{\natexlab{a}}.

\bibitem[{Di Paola} et~al.(2012{\natexlab{b}}){Di Paola}, Paci, Santoni, {De
  Ruvo}, and Giuliani]{di2012proteins}
L.~{Di Paola}, P.~Paci, D.~Santoni, M.~{De Ruvo}, and A.~Giuliani.
\newblock {Proteins as sponges: a statistical journey along protein structure
  organization principles}.
\newblock \emph{Journal of chemical information and modeling}, 52\penalty0
  (2):\penalty0 474--482, 2012{\natexlab{b}}.

\bibitem[Dorogovtsev et~al.(2008)Dorogovtsev, Goltsev, and
  Mendes]{dorogovtsev2008critical}
S.~N. Dorogovtsev, A.~V. Goltsev, and J.~F.~F. Mendes.
\newblock Critical phenomena in complex networks.
\newblock \emph{Reviews of Modern Physics}, 80\penalty0 (4):\penalty0 1275,
  2008.

\bibitem[Duardo-S{\'a}nchez et~al.(2013)Duardo-S{\'a}nchez, Munteanu,
  Riera-Fern{\'a}ndez, L{\'o}pez-D{\'\i}az, Pazos, and
  Gonz{\'a}lez-D{\'i}az]{duardo2013modeling}
A.~Duardo-S{\'a}nchez, C.~R. Munteanu, P.~Riera-Fern{\'a}ndez,
  A.~L{\'o}pez-D{\'\i}az, A.~Pazos, and H.~Gonz{\'a}lez-D{\'i}az.
\newblock Modeling complex metabolic reactions, ecological systems, and
  financial and legal networks with {MIANN} models based on markov-wiener node
  descriptors.
\newblock \emph{Journal of Chemical Information and Modeling}, 54\penalty0
  (1):\penalty0 16--29, 2013.
\newblock \doi{10.1021/ci400280n}.

\bibitem[Enright and Leitner(2005)]{PhysRevE.71.011912}
M.~B. Enright and D.~M. Leitner.
\newblock {Mass fractal dimension and the compactness of proteins}.
\newblock \emph{Physical Review E}, 71:\penalty0 011912, Jan 2005.
\newblock \doi{10.1103/PhysRevE.71.011912}.

\bibitem[Escolano et~al.(2012)Escolano, Hancock, and Lozano]{escolano2012heat}
F.~Escolano, E.~R. Hancock, and M.~A. Lozano.
\newblock {Heat diffusion: Thermodynamic depth complexity of networks}.
\newblock \emph{Physical Review E}, 85\penalty0 (3):\penalty0 036206, 2012.

\bibitem[Estrada(2010)]{estrada2010universality}
E.~Estrada.
\newblock Universality in protein residue networks.
\newblock \emph{Biophysical Journal}, 98\penalty0 (5):\penalty0 890--900, 2010.

\bibitem[Frauenfelder and Wolynes(1994)]{frauenfelder1994biomolecules}
H.~Frauenfelder and P.~G. Wolynes.
\newblock Biomolecules: where the physics of complexity and simplicity meet.
\newblock \emph{Physics Today}, 47\penalty0 (2):\penalty0 58--66, 1994.

\bibitem[Gallos et~al.(2007)Gallos, Song, Havlin, and Makse]{gallos2007scaling}
L.~K. Gallos, C.~Song, S.~Havlin, and H.~A. Makse.
\newblock Scaling theory of transport in complex biological networks.
\newblock \emph{Proceedings of the National Academy of Sciences}, 104\penalty0
  (19):\penalty0 7746--7751, 2007.
\newblock \doi{10.1073/pnas.0700250104}.

\bibitem[Gallos et~al.(2012)Gallos, Makse, and Sigman]{gallos2012small}
L.~K. Gallos, H.~A. Makse, and M.~Sigman.
\newblock {A small world of weak ties provides optimal global integration of
  self-similar modules in functional brain networks}.
\newblock \emph{Proceedings of the National Academy of Sciences}, 109\penalty0
  (8):\penalty0 2825--2830, 2012.
\newblock \doi{10.1073/pnas.1106612109}.

\bibitem[Giuliani et~al.(2014)Giuliani, Filippi, and
  Bertolaso]{giuliani2014network}
A.~Giuliani, S.~Filippi, and M.~Bertolaso.
\newblock Why network approach can promote a new way of thinking in biology.
\newblock \emph{Frontiers in genetics}, 5\penalty0 (83), 2014.

\bibitem[Gonz{\'a}lez-D{\'i}az and
  Riera-Fern{\'a}ndez(2012)]{doi:10.1021/ci300321f}
H.~Gonz{\'a}lez-D{\'i}az and P.~Riera-Fern{\'a}ndez.
\newblock New markov-autocorrelation indices for re-evaluation of links in
  chemical and biological complex networks used in metabolomics, parasitology,
  neurosciences, and epidemiology.
\newblock \emph{Journal of Chemical Information and Modeling}, 52\penalty0
  (12):\penalty0 3331--3340, 2012.
\newblock \doi{10.1021/ci300321f}.

\bibitem[Granek(2011)]{granek2011proteins}
R.~Granek.
\newblock Proteins as fractals: role of the hydrodynamic interaction.
\newblock \emph{Physical Review E}, 83\penalty0 (2):\penalty0 020902, 2011.

\bibitem[Guelzim et~al.(2002)Guelzim, Bottani, Bourgine, and
  K{\'e}p{\`e}s]{guelzim2002topological}
N.~Guelzim, S.~Bottani, P.~Bourgine, and F.~K{\'e}p{\`e}s.
\newblock Topological and causal structure of the yeast transcriptional
  regulatory network.
\newblock \emph{Nature genetics}, 31\penalty0 (1):\penalty0 60--63, 2002.

\bibitem[Guimera et~al.(2004)Guimera, Sales-Pardo, and
  Amaral]{guimera2004modularity}
R.~Guimera, M.~Sales-Pardo, and L.~A.~N. Amaral.
\newblock Modularity from fluctuations in random graphs and complex networks.
\newblock \emph{Physical Review E}, 70\penalty0 (2):\penalty0 025101, 2004.

\bibitem[Gutman and Zhou(2006)]{gutman2006laplacian}
I.~Gutman and B.~Zhou.
\newblock {Laplacian energy of a graph}.
\newblock \emph{Linear Algebra and its Applications}, 414\penalty0
  (1):\penalty0 29--37, 2006.

\bibitem[Han et~al.(2012)Han, Escolano, Hancock, and Wilson]{Han20121958}
L.~Han, F.~Escolano, E.~R. Hancock, and R.~C. Wilson.
\newblock {Graph characterizations from von Neumann entropy}.
\newblock \emph{Pattern Recognition Letters}, 33\penalty0 (15):\penalty0
  1958--1967, 2012.
\newblock ISSN 0167-8655.
\newblock \doi{10.1016/j.patrec.2012.03.016}.

\bibitem[H{\"o}fling and Franosch(2013)]{0034-4885-76-4-046602}
F.~H{\"o}fling and T.~Franosch.
\newblock Anomalous transport in the crowded world of biological cells.
\newblock \emph{Reports on Progress in Physics}, 76\penalty0 (4):\penalty0
  046602, 2013.
\newblock \doi{10.1088/0034-4885/76/4/046602}.

\bibitem[Holme and Saram{\"a}ki(2012)]{holme2012temporal}
P.~Holme and J.~Saram{\"a}ki.
\newblock Temporal networks.
\newblock \emph{Physics Reports}, 519\penalty0 (3):\penalty0 97--125, 2012.

\bibitem[Imry(1997)]{imry1997introduction}
Y.~Imry.
\newblock \emph{Introduction to mesoscopic physics}.
\newblock Oxford Univ. Press, 1997.

\bibitem[Jeong et~al.(2000)Jeong, Tombor, Albert, Oltvai, and
  Barab{\'a}si]{jeong2000large}
H.~Jeong, B.~Tombor, R.~Albert, Z.~N. Oltvai, and A.-L. Barab{\'a}si.
\newblock The large-scale organization of metabolic networks.
\newblock \emph{Nature}, 407\penalty0 (6804):\penalty0 651--654, 2000.

\bibitem[Kwapie{\'n} and Dro{\.z}d{\.z}(2012)]{kwapien2012physical}
J.~Kwapie{\'n} and S.~Dro{\.z}d{\.z}.
\newblock Physical approach to complex systems.
\newblock \emph{Physics Reports}, 515\penalty0 (3):\penalty0 115--226, 2012.
\newblock \doi{10.1016/j.physrep.2012.01.007}.

\bibitem[Laughlin et~al.(2000)Laughlin, Pines, Schmalian, Stojkovi{\'c}, and
  Wolynes]{laughlin2000middle}
R.~B. Laughlin, D.~Pines, J.~Schmalian, B.~P. Stojkovi{\'c}, and P.~Wolynes.
\newblock The middle way.
\newblock \emph{Proceedings of the National Academy of Sciences}, 97\penalty0
  (1):\penalty0 32--37, 2000.

\bibitem[Leitner(2008)]{doi:10.1146/annurev.physchem.59.032607.093606}
D.~M. Leitner.
\newblock {Energy Flow in Proteins}.
\newblock \emph{Annual Review of Physical Chemistry}, 59\penalty0 (1):\penalty0
  233--259, 2008.
\newblock \doi{10.1146/annurev.physchem.59.032607.093606}.
\newblock PMID: 18393676.

\bibitem[Lervik et~al.(2010)Lervik, Bresme, Kjelstrup, Bedeaux, and
  Rubi]{lervik2010heat}
A.~Lervik, F.~Bresme, S.~Kjelstrup, D.~Bedeaux, and J.~M. Rubi.
\newblock Heat transfer in protein--water interfaces.
\newblock \emph{Physical Chemistry Chemical Physics}, 12\penalty0 (7):\penalty0
  1610--1617, 2010.

\bibitem[Li et~al.(2014)Li, Magana, and Dyer]{li2014anisotropic}
G.~Li, D.~Magana, and R.~B. Dyer.
\newblock Anisotropic energy flow and allosteric ligand binding in albumin.
\newblock \emph{Nature communications}, 5, 2014.

\bibitem[Livi and Rizzi(2013)]{Livi_ga_2013}
L.~Livi and A.~Rizzi.
\newblock {Graph ambiguity}.
\newblock \emph{Fuzzy Sets and Systems}, 221:\penalty0 24--47, 2013.
\newblock ISSN 0165-0114.
\newblock \doi{10.1016/j.fss.2013.01.001}.

\bibitem[Livi et~al.(2014{\natexlab{a}})Livi, Giuliani, and
  Rizzi]{ecoli_graph__arxiv}
L.~Livi, A.~Giuliani, and A.~Rizzi.
\newblock Toward a multilevel representation of protein molecules: comparative
  approaches to the aggregation/folding propensity problem.
\newblock \emph{ArXiv preprint arXiv:1407.7559}, Jul 2014{\natexlab{a}}.

\bibitem[Livi et~al.(2014{\natexlab{b}})Livi, Giuliani, and
  Sadeghian]{ecoli_graph_complexity_arxiv}
L.~Livi, A.~Giuliani, and A.~Sadeghian.
\newblock Characterization of graphs for protein structure modeling and
  recognition of solubility.
\newblock \emph{arXiv preprint arXiv:1407.8033}, Jul 2014{\natexlab{b}}.

\bibitem[Mirshahvalad et~al.(2014)Mirshahvalad, Esquivel, Lizana, and
  Rosvall]{mirshahvalad2014dynamics}
A.~Mirshahvalad, A.~V. Esquivel, L.~Lizana, and M.~Rosvall.
\newblock Dynamics of interacting information waves in networks.
\newblock \emph{Physical Review E}, 89\penalty0 (1):\penalty0 012809, 2014.

\bibitem[Mitrovi{\'c} and Tadi{\'c}(2009)]{mitrovic2009spectral}
M.~Mitrovi{\'c} and B.~Tadi{\'c}.
\newblock Spectral and dynamical properties in classes of sparse networks with
  mesoscopic inhomogeneities.
\newblock \emph{Physical Review E}, 80\penalty0 (2):\penalty0 026123, 2009.

\bibitem[Morita and Takano(2009)]{PhysRevE.79.020901}
H.~Morita and M.~Takano.
\newblock Residue network in protein native structure belongs to the
  universality class of a three-dimensional critical percolation cluster.
\newblock \emph{Physical Review E}, 79:\penalty0 020901, Feb 2009.
\newblock \doi{10.1103/PhysRevE.79.020901}.

\bibitem[Nakayama et~al.(1994)Nakayama, Yakubo, and
  Orbach]{nakayama1994dynamical}
T.~Nakayama, K.~Yakubo, and R.~L. Orbach.
\newblock Dynamical properties of fractal networks: Scaling, numerical
  simulations, and physical realizations.
\newblock \emph{Reviews of Modern Physics}, 66\penalty0 (2):\penalty0 381,
  1994.

\bibitem[Neusius et~al.(2008)Neusius, Daidone, Sokolov, and
  Smith]{neusius2008subdiffusion}
T.~Neusius, I.~Daidone, I.~M. Sokolov, and J.~C. Smith.
\newblock Subdiffusion in peptides originates from the fractal-like structure
  of configuration space.
\newblock \emph{Physical Review Letters}, 100\penalty0 (18):\penalty0 188103,
  2008.

\bibitem[Newman(2005)]{newman2005power}
M.~E.~J. Newman.
\newblock Power laws, pareto distributions and zipf's law.
\newblock \emph{Contemporary Physics}, 46\penalty0 (5):\penalty0 323--351,
  2005.

\bibitem[Newman(2006)]{newman2006modularity}
M.~E.~J. Newman.
\newblock Modularity and community structure in networks.
\newblock \emph{Proceedings of the National Academy of Sciences}, 103\penalty0
  (23):\penalty0 8577--8582, 2006.

\bibitem[Newman(2010)]{newman2010networks}
M.~E.~J. Newman.
\newblock \emph{Networks: an introduction}.
\newblock Oxford University Press, 2010.

\bibitem[Nicosia et~al.(2014)Nicosia, Domenico, and
  Latora]{nicosia2013characteristic}
V.~Nicosia, M.~D. Domenico, and V.~Latora.
\newblock Characteristic exponents of complex networks.
\newblock \emph{EPL (Europhysics Letters)}, 106\penalty0 (5):\penalty0 58005,
  2014.

\bibitem[Niwa et~al.(2009)Niwa, Ying, Saito, Jin, Takada, Ueda, and
  Taguchi]{niwa2009}
T.~Niwa, B.-W. Ying, K.~Saito, W.~Jin, S.~Takada, T.~Ueda, and H.~Taguchi.
\newblock {Bimodal protein solubility distribution revealed by an aggregation
  analysis of the entire ensemble of Escherichia coli proteins}.
\newblock \emph{Proceedings of the National Academy of Sciences}, 106\penalty0
  (11):\penalty0 4201--4206, 2009.
\newblock \doi{10.1073/pnas.0811922106}.

\bibitem[Orozco(2014)]{orozco2014theoretical}
M.~Orozco.
\newblock A theoretical view of protein dynamics.
\newblock \emph{Chemical Society Reviews}, 43:\penalty0 5051--5066, 2014.
\newblock \doi{10.1039/C3CS60474H}.

\bibitem[Pinna et~al.(2010)Pinna, Soranzo, and {De La
  Fuente}]{pinna2010knockouts}
A.~Pinna, N.~Soranzo, and A.~{De La Fuente}.
\newblock From knockouts to networks: establishing direct cause-effect
  relationships through graph analysis.
\newblock \emph{PloS one}, 5\penalty0 (10):\penalty0 e12912, 2010.

\bibitem[Pinna et~al.(2011)Pinna, Soranzo, Hoeschele, and {de la
  Fuente}]{pinna2011simulating}
A.~Pinna, N.~Soranzo, I.~Hoeschele, and A.~{de la Fuente}.
\newblock Simulating systems genetics data with sysgensim.
\newblock \emph{Bioinformatics}, 27\penalty0 (17):\penalty0 2459--2462, 2011.

\bibitem[Reuveni et~al.(2010)Reuveni, Granek, and
  Klafter]{reuveni2010anomalies}
S.~Reuveni, R.~Granek, and J.~Klafter.
\newblock Anomalies in the vibrational dynamics of proteins are a consequence
  of fractal-like structure.
\newblock \emph{Proceedings of the National Academy of Sciences}, 107\penalty0
  (31):\penalty0 13696--13700, 2010.

\bibitem[Reuveni et~al.(2012)Reuveni, Klafter, and Granek]{reuveni2012dynamic}
S.~Reuveni, J.~Klafter, and R.~Granek.
\newblock Dynamic structure factor of vibrating fractals: Proteins as a case
  study.
\newblock \emph{Physical Review E}, 85\penalty0 (1):\penalty0 011906, 2012.

\bibitem[Riera-Fernandez et~al.(2012)Riera-Fernandez, Munteanu, Escobar,
  Prado-Prado, Mart{\'\i}n-Romalde, Pereira, Villalba, Duardo-Sanchez, and
  Gonz{\'a}lez-D{\'\i}az]{riera2012new}
P.~Riera-Fernandez, C.~R. Munteanu, M.~Escobar, F.~Prado-Prado,
  R.~Mart{\'\i}n-Romalde, D.~Pereira, K.~Villalba, A.~Duardo-Sanchez, and
  H.~Gonz{\'a}lez-D{\'\i}az.
\newblock New {M}arkov--{S}hannon entropy models to assess connectivity quality
  in complex networks: from molecular to cellular pathway, parasite--host,
  neural, industry, and legal--social networks.
\newblock \emph{Journal of Theoretical Biology}, 293:\penalty0 174--188, 2012.
\newblock \doi{10.1016/j.jtbi.2011.10.016}.

\bibitem[Rossi et~al.(2013)Rossi, Torsello, Hancock, and
  Wilson]{rossi2013characterizing}
L.~Rossi, A.~Torsello, E.~R. Hancock, and R.~C. Wilson.
\newblock Characterizing graph symmetries through quantum jensen-shannon
  divergence.
\newblock \emph{Physical Review E}, 88\penalty0 (3):\penalty0 032806, 2013.

\bibitem[Rosvall et~al.(2014)Rosvall, Esquivel, Lancichinetti, West, and
  Lambiotte]{rosvall2014memory}
M.~Rosvall, A.~V. Esquivel, A.~Lancichinetti, J.~D. West, and R.~Lambiotte.
\newblock Memory in network flows and its effects on spreading dynamics and
  community detection.
\newblock \emph{Nature Communications}, 5, 2014.

\bibitem[Sangha and Keyes(2009)]{sangha2009proteins}
A.~K. Sangha and T.~Keyes.
\newblock Proteins fold by subdiffusion of the order parameter.
\newblock \emph{The Journal of Physical Chemistry B}, 113\penalty0
  (48):\penalty0 15886--15894, 2009.

\bibitem[Song et~al.(2005)Song, Havlin, and Makse]{song2005self}
C.~Song, S.~Havlin, and H.~A. Makse.
\newblock {Self-similarity of complex networks}.
\newblock \emph{Nature}, 433\penalty0 (7024):\penalty0 392--395, 2005.

\bibitem[Song et~al.(2006)Song, Havlin, and Makse]{song2006origins}
C.~Song, S.~Havlin, and H.~A. Makse.
\newblock Origins of fractality in the growth of complex networks.
\newblock \emph{Nature Physics}, 2\penalty0 (4):\penalty0 275--281, 2006.

\bibitem[Stumpf et~al.(2008)Stumpf, Thorne, {de Silva}, Stewart, An, Lappe, and
  Wiuf]{stumpf2008estimating}
M.~P.~H. Stumpf, T.~Thorne, E.~{de Silva}, R.~Stewart, H.~J. An, M.~Lappe, and
  C.~Wiuf.
\newblock Estimating the size of the human interactome.
\newblock \emph{Proceedings of the National Academy of Sciences}, 105\penalty0
  (19):\penalty0 6959--6964, 2008.

\bibitem[Trinajsti{\'c}(1983)]{trinajstiac1983chemical}
N.~Trinajsti{\'c}.
\newblock \emph{{Chemical graph theory}}.
\newblock CRC Press, Boca Raton, FL, 1983.

\bibitem[Tsallis(2001)]{tsallis2001nonextensive}
C.~Tsallis.
\newblock I. nonextensive statistical mechanics and thermodynamics: Historical
  background and present status.
\newblock In \emph{Nonextensive statistical mechanics and its applications},
  pages 3--98. Springer, 2001.

\bibitem[Vijayabaskar and Vishveshwara(2010)]{vijayabaskar2010interaction}
M.~S. Vijayabaskar and S.~Vishveshwara.
\newblock Interaction energy based protein structure networks.
\newblock \emph{Biophysical Journal}, 99\penalty0 (11):\penalty0 3704--3715,
  2010.

\bibitem[Weaver(1991)]{weaver1991science}
W.~Weaver.
\newblock Science and complexity.
\newblock In \emph{Facets of Systems Science}, pages 449--456. Springer, 1991.

\bibitem[Xiao et~al.(2009)Xiao, Hancock, and
  Wilson]{Xiao:2009:GCH:1563046.1563099}
B.~Xiao, E.~R. Hancock, and R.~C. Wilson.
\newblock {Graph Characteristics from the Heat Kernel Trace}.
\newblock \emph{Pattern Recognition}, 42\penalty0 (11):\penalty0 2589--2606,
  Nov. 2009.
\newblock ISSN 0031-3203.
\newblock \doi{10.1016/j.patcog.2008.12.029}.

\bibitem[Yan et~al.(2014)Yan, Zhou, Sun, Chen, Hu, and
  Shen]{yan2014construction}
W.~Yan, J.~Zhou, M.~Sun, J.~Chen, G.~Hu, and B.~Shen.
\newblock The construction of an amino acid network for understanding protein
  structure and function.
\newblock \emph{Amino Acids}, 46\penalty0 (6):\penalty0 1419--1439, 2014.

\bibitem[Yu and Leitner(2003)]{yu2003anomalous}
X.~Yu and D.~M. Leitner.
\newblock Anomalous diffusion of vibrational energy in proteins.
\newblock \emph{The Journal of Chemical Physics}, 119\penalty0 (23):\penalty0
  12673--12679, 2003.

\end{thebibliography}
\end{document}